\documentclass[12pt,epsf]{article}
  \usepackage{epsfig}
  \newlength{\abstractwidth}
  \renewcommand{\thefootnote}{\fnsymbol{footnote}}
  \renewcommand{\thanks}[1]{\footnote{#1}} % Use this for footnotes
  \newcommand{\starttext}{
  \setcounter{footnote}{0}
  \renewcommand{\thefootnote}{\arabic{footnote}}}
  \renewcommand{\theequation}{\thesection.\arabic{equation}}
  \newcommand{\be}{\begin{equation}}
  \newcommand{\bea}{\begin{eqnarray}}
  \newcommand{\eea}{\end{eqnarray}}
  \newcommand{\beq}{\begin{equation}}
  \newcommand{\ee}{\end{equation}}
  \newcommand{\eeq}{\end{equation}}
  
  \newcommand{\<}{\langle\,}

  \renewcommand{\>}{\rangle}

  \def\ba{\begin{eqnarray}}
  \def\ea{\end{eqnarray}}

\def\wdw{Wheeler-DeWitt}

  \def\12{{1 \over 2}}
  \def\eq{&=&}

  \def\h3{h^{3\over 2}}

  \def\cc{cosmological constant }
  
  \def\simleq{\; \raise0.3ex\hbox{$<$\kern-0.75em
      \raise-1.1ex\hbox{$\sim$}}\; }
   \def\simgeq{\; \raise0.3ex\hbox{$>$\kern-0.75em
      \raise-1.1ex\hbox{$\sim$}}\; }

\def\sig{$\Sigma$}
\def\ads{anti de Sitter space}
\def\cft{conformal field theory}
\def\cdl{Coleman De Luccia}

\def\ba{\bf{a}}

\def\ic{information-capacity}
  
\begin{document}
  \renewcommand{\theequation}{\thesection.\arabic{equation}}
  \begin{titlepage}

\bigskip\bigskip
\rightline{SU-ITP-09/40}
\rightline{OIQP-09-09}

  \bigskip\bigskip\bigskip\bigskip
  \bigskip \bigskip

  \centerline{\Large \bf { Census Taking in the Hat:}}
    \centerline{\Large \bf { FRW/CFT Duality}}

  \bigskip \bigskip

  \bigskip\bigskip
  \bigskip\bigskip
  %\centerline{\it }
  %\medskip
  %\centerline{} \centerline{} \centerline{}
  %\bigskip

\begin{center}
  {{\large Yasuhiro Sekino${}^{1,2}$ and Leonard Susskind${}^{1}$ }}
  \bigskip

\bigskip
${}^1\;$   Department of Physics, Stanford University\\ 
Stanford, CA 94305-4060, USA \\ 

\medskip

${}^2\;$ Okayama Institute for Quantum Physics\\ 
1-9-1 Kyoyama, Okayama 700-0015, Japan
\vspace{2cm}
  \end{center}

  \bigskip\bigskip
  \begin{abstract}

% Abstract
In this paper a holographic description of eternal inflation is developed.  We focus on the description of an open FRW universe that results from a tunneling event in which a false  vacuum with positive vacuum energy decays to a supersymmetric vacuum with vanishing cosmological constant. The observations of a ``Census Taker" in the final vacuum can be organized into a holographic dual conformal field theory that lives on the asymptotic boundary of space.  We refer to this bulk-boundary correspondence as FRW/CFT duality. The dual CFT is a Euclidean two-dimensional theory that includes a Liouville 2-D gravity sector describing geometric fluctuations of the boundary. The RG flow of the theory is richer than in the ADS/CFT correspondence, and generates two space-time dimensions---one space-like and one time-like. We discuss a number of phenomena such as bubble collisions, and the Garriga, Guth Vilenkin ``persistence of memory,"  from the dual viewpoint.

  \medskip
  \noindent
  \end{abstract}

  \end{titlepage}
  \starttext \baselineskip=17.63pt \setcounter{footnote}{0}

%%%%%%%%%%%%%%%%
%%%%%%%%%%%%%%%%
%%%%%%%%%%%%%%%%
%%%%%%%%%%%%%%%%
%%%%%%%%%%%%%%%%
%%%%%%%%%%%%%%%%

\bigskip

\setcounter{equation}{0}
  \section{ Introduction }

There are two views of eternal inflation\cite{eternal}. According to the  ``global" point of view  the   entire multiverse  is an infinite system of pocket universes populating a Landscape \cite{bousso-polch,kklt,landscape,douglas}. The global view raises some difficult questions of principle: for example, in saying the multiverse is a \it system, \rm do we mean  a quantum-system described by  a wave function $\Psi$?; if so what variables does $\Psi$ depend on?; how local is the description?; and what relevance does the existence of cosmic event-horizons have? If the description is (approximately) local then what operational meaning can be attached to correlation functions between variables in regions that are out of causal contact?

According to the second   local or causal patch viewpoint \cite{entropic} cosmology should be formulated without any reference to unobservable events beyond the observer's horizon. In such a formulation the wave function $\Psi$ depends only on the degrees of  freedom that have operational meaning to the observer. The biggest question raised by the local viewpoint is what meaning to attach to events beyond the observer's horizon.

In this paper we will focus on  the local description of  an observer located in a so-called terminal vacuum with exactly vanishing cosmological constant. Following Shenker, we will call such an observer a Census Taker (CT). However, we will also give reasons to believe that  the global and the local descriptions are both  correct,  being related by a complementarity principle similar to  black hole complementarity.

Obviously the causal past of the Census Taker contains the most information in the  asymptotic late-time limit. Thus, to define the  CT's space of states we need to specify the late-time limit of the CT's trajectory. All trajectories in an eternally inflating universe eventually end in some kind of terminal state. The terminal states are easiest to describe if we assume that the mechanism for de Sitter decay is
 bubble nucleation of the type considered by Coleman and DeLuccia \cite{Coleman}. The result of such a decay is always
 an
open, negatively curved, FRW universe. This paper will expand on  the  proposal of \cite{yeh}, in which a holographic duality was introduced between  an FRW cosmology, and a two-dimensional Euclidean conformal field theory. This duality will be called the FRW/CFT correspondence. The regulator that cuts off the Census Taker's observations is time: the longer the CT observes, the more he counts.  Using the dictionary provided by FRW/CFT, the cutoff may be identified with the ultraviolet regulator of the 2-dimensional CFT.

\setcounter{equation}{0}
  \section{The Census Bureau}

  Let us begin with a precise definition of a causal patch. Start with a cosmological
space-time and assume that a future causal boundary exists. For
example, in flat Minkowski space the future causal boundary
consists of ${\cal{I}}^+$ (future light like infinity) and a
single point: time-like-infinity.
  For a  Schwarzschild black hole, the future causal boundary has an
additional component: the  singularity.

  A causal patch is defined in terms of a point $\bf\underline{a}$ on the future causal
boundary. Call that point the ``Census Bureau\footnote{This
term originated during a discussion with Steve
Shenker.}."  By
definition, the causal patch is  the causal past of the Census Bureau,  bounded by its
past light cone. For Minkowski space, one usually  picks the
Census Bureau to be time-like-infinity. In that case the causal
patch is all of Minkowski space as seen in figure 1.

 \begin{figure}
\begin{center}
\includegraphics[width=12cm]{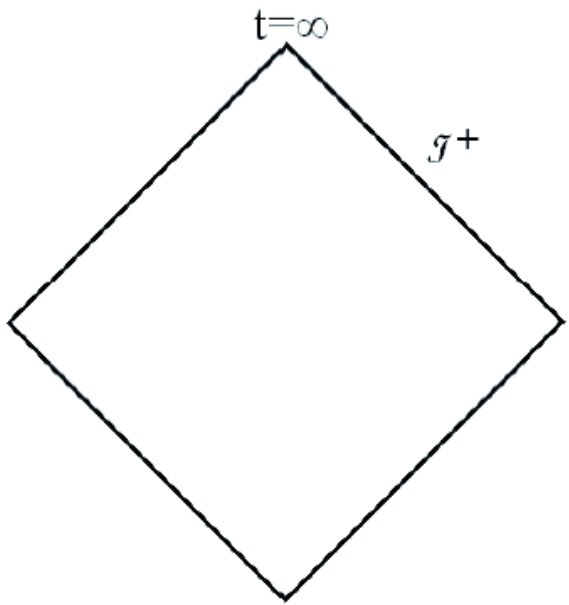}
\caption{Conformal diagram for ordinary flat Minkowski space. The
 causal patch associated with the ``Census Bureau" at $t=\infty$ is the entire space-time. } \label{1}
\end{center}
\end{figure}

  In the case of the Schwarzschild geometry, $\bf\underline{a}$ can again be chosen to be
time-like-infinity, in which case the causal patch is everything
outside the horizon of the black hole. There is no clear reason
why one can't choose $\bf\underline{a}$ to be on the singularity\cite{gavin},
but it  would lead to obvious difficulties.

de Sitter space has the causal structure  shown in
figure 2. In this case all  points at  future infinity are
equivalent: the Census Bureau can be located at any of them.
However, String Theory and other considerations \cite{trouble} suggest that  de
Sitter minima are never stable. After a series of tunneling events
they eventually end in
 terminal vacua with exactly zero or negative cosmological
constant. The entire distant future of de Sitter space is replaced
by a fractal of terminal bubbles.

\begin{figure}
\begin{center}
\includegraphics[width=12cm]{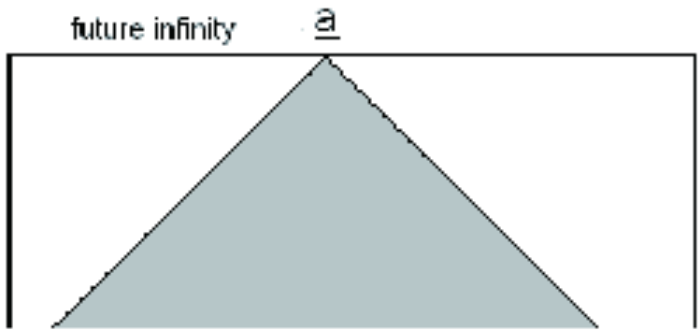}
\caption{Conformal diagrams for eternal  de Sitter
space.  The causal past of the Census Bureau at 
 $\bf{a}$ is shown in grey. } \label{2}
\end{center}
\end{figure}

Transitions to vacua with negative cosmological constant always lead to  singular crunches in which the energy density blows up, or approaches the Planck scale.  As in the case of the black hole, we will not consider Census Bureaus located on singularities.  That leaves only the supersymmetric bubbles with zero cosmological constant\footnote{We will assume that non-supersymmetric vacua never have exactly zero cosmological constant. }.
Such a bubble  evolves to  an open, negatively curved, FRW universe  bounded
by a ``hat" \cite{yeh}. The Census Bureau is at the tip of the
hat.

The term ``Census Taker"  denotes
an observer in such an FRW universe who looks back into
the past and collects data. He can count galaxies, other
observers, hydrogen atoms, colliding bubble-universes,
civilizations, or anything else within his own causal past. As
time elapses the Census Taker sees more and more of the causal
patch. Eventually all  Census takers within the same causal patch
arrive at the Census Bureau where they can compare data.

There are two possible connections  between hatted terminal geometries, and
observational cosmology with a non-zero cosmological constant.
First,  for many purposes, the
current \cc  is so small that  it can be set to zero. Later we
will argue that the \cft \ description of the approximate hat
which results from non-zero cosmological constant is an
ultraviolet incomplete version of the type of field theory that
describes a hat.

 This paper is  mainly concerned with the second connection in which hatted geometries are used as probes of eternal inflation. It was emphasized by Shenker, et al. \cite{shenker} \  that because
any de Sitter vacuum will eventually decay, a
Census Taker can look back into it from a point at
or near the tip of a hat, and gather information. In principle the
Census Taker can look back, not only into the Ancestor vacuum (our
vacuum in this case), but also into bubble collisions with other
vacua of the Landscape. Much of this paper is about the gathering
of information as the Census Taker's time progresses, and how it
is encoded in the renormalization-group (RG) flow of a holographic
field theory.

\subsection{Asymptotic Coldness}
String Theory is  a  powerful tool in the study of quantum gravity, but only in
 special  backgrounds such as flat space and ADS space.
Effective as it is in describing
scattering amplitudes in asymptotically flat (supersymmetric) space-time, and
black holes in \ads,  it is an inflexible tool which at present is
not useful as a  mathematical  framework for
cosmology. What is it that is so special about flat and \ads \ backgrounds
that allows a rigorous formulation of quantum gravity, and why are
cosmological backgrounds so difficult?

The problem is  frequently blamed on
time-dependence, but time-dependent deformations
of \ads \ or Matrix Theory \cite{bfss} are easy to describe.
Time dependence in itself does not seem to be the problem.
There is one important difference between the usual
String Theory backgrounds and  more interesting  cosmological
backgrounds. Asymptotically-flat and anti de Sitter backgrounds
have a property that we will call \it asymptotic
coldness.\rm \  Asymptotic coldness  means  that the boundary
conditions require the energy density to go to zero at the
asymptotic boundary of space-time. Similarly, the fluctuations in geometry
tend to zero.  This condition is embodied in the statement that
all physical disturbances are composed  of normalizable modes.
Asymptotic coldness is obviously important to defining an S-matrix
in flat space-time, and plays an equally important role in
defining the observables of \ads.

But in cosmology, asymptotic coldness is  never the case. Closed
universes have no asymptotic boundary,  and homogeneous infinite
universes have matter, energy, and geometric variation out to
spatial infinity. Under this circumstance an S-matrix cannot be
formulated. String theory at present is ill equipped to deal with
asymptotically warm geometries. To put it another
way, there is a conflict between a homogeneous cosmology, and the
Holographic Principle \cite{hologram,juan,witten,susskind and witten} which requires an isolated, cold, boundary.

Consider the three kinds of decay products that can occur in eternally inflating space-time: de Sitter bubbles with positive cosmological constant: crunching bubbles with negative cosmological constant: and supersymmetric hatted geometries with zero cosmological constant. de Sitter geometries are never asymptotically cold; the thermal fluctuations continue forever or at least until the de Sitter space decays. Crunches are obviously not good candidates for asymptotic coldness. That leaves  vacua with vanishing cosmological constant bounded by hats.

The geometry under a hat is not asymptotically cold.  A negatively curved FRW universe is spatially homogeneous on large scales, and not empty. Thus if we fix the time and go out to spatial infinity conditions do not become cold.
However, if the cosmological constant is zero, the temperature and density of matter do tend to zero at asymptotically late times. Although we may not be able to define an S-matrix, the late time universe can be described in terms of a supersymmetric spectrum of free particles described by string theory. This partial asymptotic coldness makes hatted geometries  the best candidates for a precise mathematical formulation of eternally inflating cosmology.

\setcounter{equation}{0}
  \section{The FRW/CFT Duality }

The classical space-time in the interior of a \cdl \ bubble \cite{Coleman},  has
the form of an open infinite FRW universe. Let  ${\cal{H}}_3$
represent a hyperbolic geometry with constant negative curvature,
\be d {\cal{H}}_3^2 = dR^2 + \sinh^2{R} \ \ d\Omega_2^2.
\label{hyperbolic} \ee The metric of open FRW is \be ds^2 = -dt^2
+ a(t)^2 d{\cal{H}}_3^2, \label{FRW metric in terms of t} \ee or
in terms of  conformal time $T$ (defined by $dT = dt/ a(t) $) \be
ds^2 = a(T)^2 ( -dT^2 + d{\cal{H}}_3^2). \label{FRW metric in terms
of T} \ee Note that in (\ref{hyperbolic}) the radial coordinate
$R$ is a dimensionless hyperbolic angle and that the symmetry of the spatial
sections is the non-compact group $O(3,1)$.  This
symmetry  plays  a central  role in what follows.

If the vacuum energy in the bubble is zero, i.e., no cosmological
constant, then the future boundary of the FRW region is a hat. The
scale factor $a(t) $ then has the  early and late-time behaviors
\bea
a(T) \sim t \sim   H^{-1}e^{(T+T_0)},
\label{asymptotic a}
\eea
where $H$ is the Hubble constant of the ancestor vacuum.
For early time when $T \to -\infty$ the constant $T_0$ is
zero, \be a(T) = H^{-1} e^T \ \ \  \ \  (T \to - \infty).
\ee 
%At late time $T_0$ is positive.
In  the  simplest thin-wall case $T_0$ is zero for all time 
(within the FRW region). In general the sign of $T_0$ at 
late time depends on the equation of state at the intermediate 
stage. If there is an accelerating (decelerating) phase between 
the early and the late time phases (\ref{asymptotic a}), 
$T_0$ is positive (negative) at late time.

\begin{figure}
\begin{center}
\includegraphics[width=12cm]{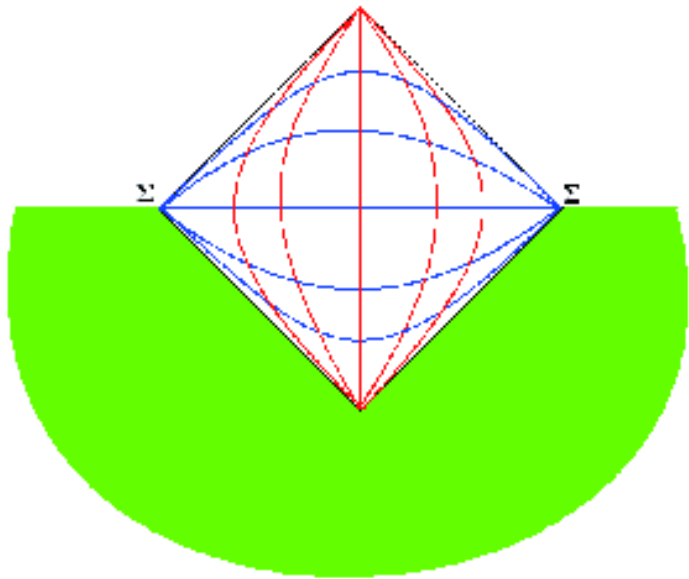}
\caption{A Conformal diagram for the FRW universe created by
bubble nucleation from an ``Ancestor" metastable vacuum. The
Ancestor vacuum is shown in green. The red and blue curves are
surfaces of constant $T$ and $R$. The two-sphere at spatial
infinity is indicated by \sig.} \label{3}
\end{center}
\end{figure}

  In figure 3,
a conformal diagram of FRW is illustrated,  with surfaces of
constant $T$ and
 $R$  shown in red and blue. The green region
represents the de Sitter Ancestor vacuum. Figure 4 shows the
Census Taker, as he approaches the tip of the hat, looking back
along his past light cone.

 \begin{figure}
\begin{center}
\includegraphics[width=12cm]{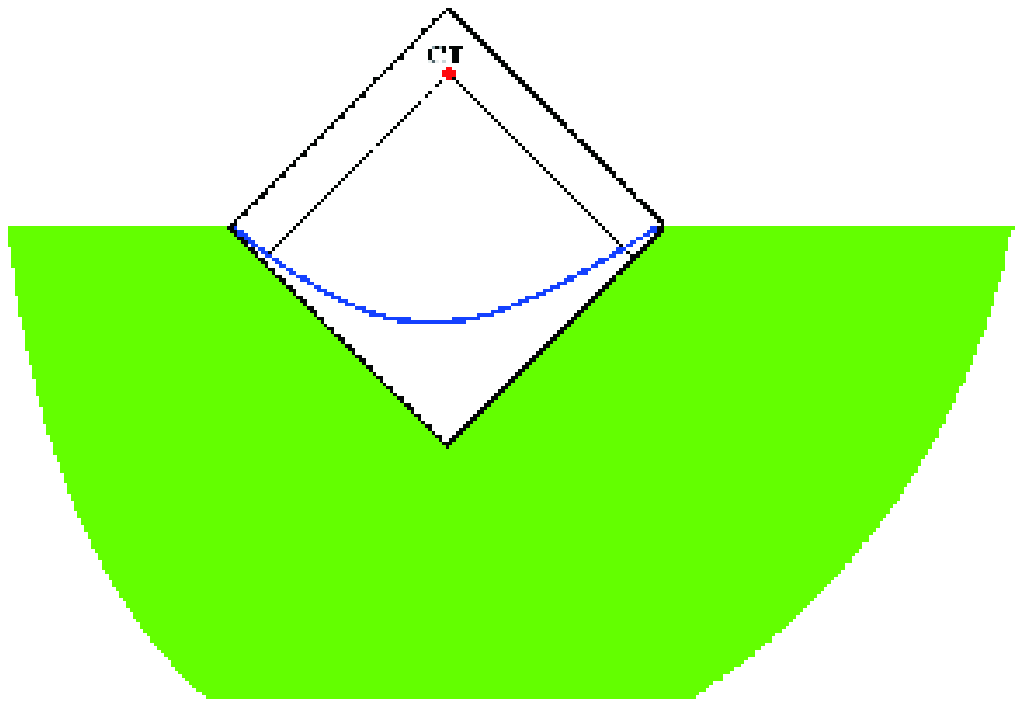}
\caption{The Census Taker is indicated by the red dot. The thin black
lines represent his past light-cone and the blue curve is a space-like  surface of constant $T$. } \label{4}
\end{center}
\end{figure}

The geometry of a spatial
slice of constant $T$ is a three
dimensional, negatively curved, hyperbolic plane. It is identical to
3-D Euclidean anti de Sitter space. The two dimensional analog is well
illustrated in Figure 5 by  Escher's  drawing ``Limit Circle IV." It is both a
drawing of Euclidean ADS and also a fixed-time slice of open FRW.

 In Figure 5,
the green circle is the intersection of Census Taker's past light
cone with the time-slice. As the Census Taker advances in time,
the green circle moves out toward the boundary.

\begin{figure}
\begin{center}
\includegraphics[width=20cm]{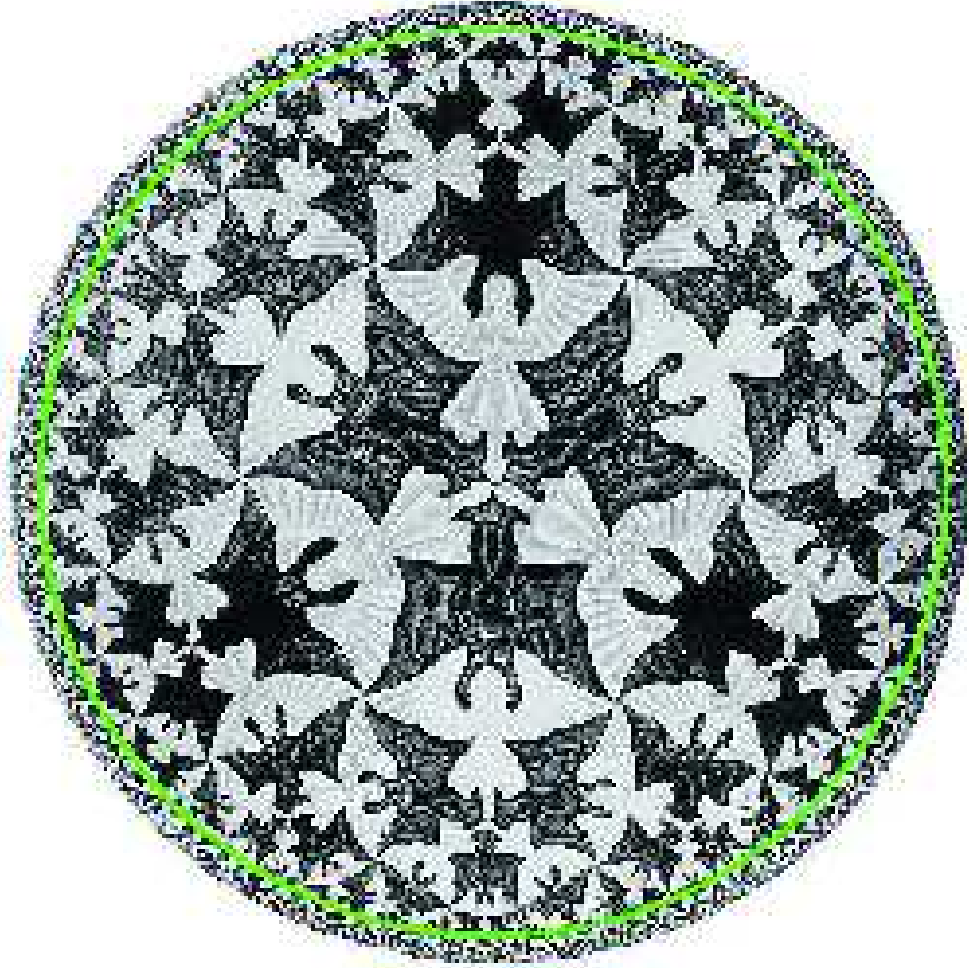}
\caption{Escher's drawing of the Hyperbolic Plane, which
represents Euclidean \ads \ or a spatial slice of open FRW. The
green circle shows the intersection of the Census Taker's past
light-cone, which moves toward the boundary with
Census-Taker-time.} \label{5}
\end{center}
\end{figure}
 \bigskip

A fact (to be explained later) which will play a leading role in
what follows, concerns the Census Taker's angular resolution,
i.e., his ability to discern small angular variation. If the time
at which the CT looks back is called $T_{CT} $, then the smallest
angle he can resolve  is of order $\exp{(-T_{CT})}$. It is as if
the CT were looking deeper and deeper into the ultraviolet
structure of a quantum field theory on \sig. This  observation
 motivates an FRW/CFT correspondence.

The boundary of \ads \ plays a key role in the ADS/CFT
correspondence, where it represents the extreme ultraviolet
degrees of freedom of the boundary theory. The corresponding
boundary in the FRW geometry is labeled \sig \ and consists of the
intersection of the hat ${\cal{I}}^+$, with the space-like future
boundary of de Sitter space. From within the interior of the
bubble, \sig  \ represents  space-like infinity.
It is the obvious surface for a holographic description. As one
might expect, the $O(3,1)$ symmetry which acts on the time-slices,
also has the action of two dimensional conformal transformations
on \sig.  Whatever the Census Taker sees, it is very natural for
him to classify his observations under the conformal group. Thus,
the apparatus of (Euclidean) conformal field theory, such as
operator dimensions, and correlation functions, should play a
leading role in organizing his data. The conjecture of \cite{yeh}
is that there exists an exact duality between the bulk description
of the hatted FRW universe and a conformal field theory living on
$\Sigma.$

In complicated situations, such as multiple bubble collisions,
\sig \ requires a precise definition. The asymptotic light-cone
${\cal{I}}^+$ (which is, of course, the limit of the Census Taker's
past light cone), can be thought of as being formed from a
collection of light-like generators.  Each generator, at one end,
runs into the tip of the hat, while the other end eventually
enters the bulk space-time. The set of points where the generators
enter the bulk define \sig.

\subsection{Observer  Complementarity and The Census Taker}

This paper is about a duality between the FRW patch under a hat, and two-dimensional conformal field theory. However, it is possible that there is a larger point at stake---a possible complementarity between the Census Taker's patch and the entire multiverse.
The question is  whether or not degrees of freedom beyond the Census Taker's  horizon have meaning (We believe the answer is yes.); and whether they are independent of the degrees of freedom  on the observer's side (In our opinion, no.). In the context of black holes the situation is fairly clear by now and is encapsulated in the principle of black hole complementarity. A generalization of black hole complementarity is sometimes called observer complementarity. In this section we will review the complementarity principle \cite{complementarity} and then discuss its possible  application to the relation between the Census Taker and the multiverse.

Consider any ordinary quantum system with a Hilbert space $\cal{H}$ and a collection of observables (We work in the Heisenberg picture.). Pick two times, one in the past and one in the future. Call them $t_{in}$ and $t_{out}$. For every observable $A(t_{out})$ (to be measured at  $t_{out}$), there is another observable that can be measured at $t_{in}$ with exactly the same spectrum, and the same probability distribution as $A(t_{out})$ in any state. To find that operator all we have to do is solve the Heisenberg equations of motion, and express  $A(t_{out})$ as a functional of the operators at $t_{in}$. An example from elementary quantum mechanics is the free particle on a line. Choose $A(t_{out})$ to be the position of the particle $X(t_{out})$. The corresponding operator at time $t_{in} $ is
$$
X(t_{in}) + {P(t_{in}) \over M }(t_{out}- t_{in})
$$

The point is not that measuring $
X(t_{in}) + {P(t_{in}) \over M }(t_{out}- t_{in})
$
at time $t_{in}$ is the same thing as measuring  $X(t_{out})$  at the later time---it is clearly not---but that
 in any Heisenberg state, the probability distribution for the two measurements are the same.

 To say it another way, imagine preparing a particle in the remote past in some state. We may measure $X$ at time $t_{out}$ or we may measure $X + {P \over M }(t_{out}- t_{in})$ at time $t_{in}$. The probability distributions for the two experiments are identical.

Thus, by solving the equations of motion one can express $A(t_{out})$ as
a functional $ A_*(t_{in}) $, of operators at time
$t_{in}$. ($A_*(t_{in}) $ has the same probability distribution as
$A(t_{out})$, and is the same Heisenberg operator as $A(t_{out}))$.
A formal expression for $A_*(t_{in}) $ is given by
\be
A_*(t_{in}) = U^{\dag}(t_{out}, t_{in}) A(t_{in}) U(t_{out}, t_{in})
\ee
%\be
%A_*(t_{in}) = U(t_{out}-t_{in}) A(t_{out}) U^{\dag}(t_{out}-t_{in})
%\ee
where $U(t_{out}, t_{in})$ is the usual time development operator from
$t_{in}$ to $t_{out}$.

Now let us consider a process in which a black hole forms and evaporates.  We begin by following an in-falling system in ``free floating coordinates," such as Eddington-Finkelstein or Painleve-Gullstand coordinates. In such coordinates the in-falling system
can be described by ordinary low energy physics as it crosses the horizon, at least until  it approaches the singularity. Consider a low energy observable $A(p)$ at a point $p$,  behind the horizon of the black hole (See Figure 6).
 \begin{figure}
\begin{center}
\includegraphics[width=12cm]{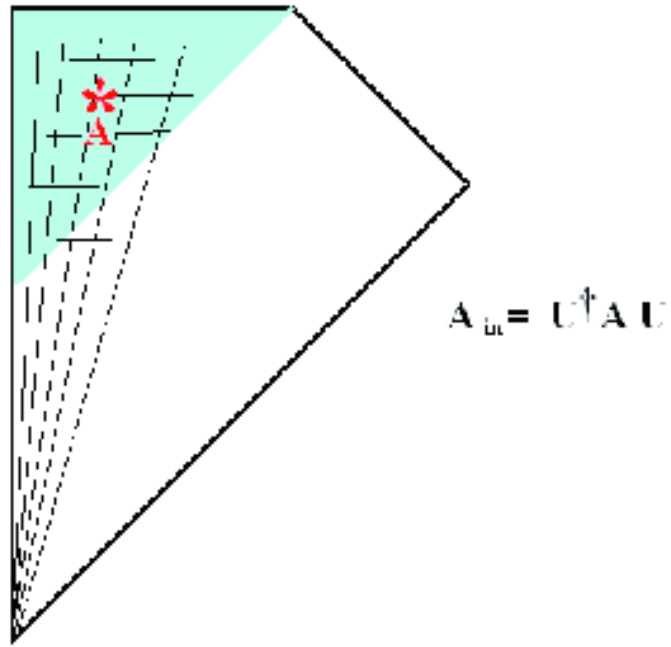}
\caption{A low energy operator $A$ behind the horizon can be evolved into the past. The useful coordinates are freely falling.} \label{event2}
\end{center}
\end{figure}
One can always find an observable $A_{*}(in)$ in the remote past,
outside the black hole,  with the same expectation value and probability
distribution as $A(p)$. 
All we need is an operator $U^{\dag}$ to run 
%All we need is low energy physics to run the
the operator backward. 
Then we get $A_{*}(in)$ by running $A(in)$ forward with $U$, 
\be
A_*({in}) = U^{\dag} A(in) U.
\ee
%where $U$ is the appropriate time development operator from the point
%$p$, back to the remote past. 
We emphasize again, that throughout this operation the description in the in-falling frame  is governed by conventional low energy  quantum field theory. Note that
$A_*({in})$ is a function of operators at time-like and light-like past infinity. In other words it is a function of asymptotic in-operators.

%Next, assume that in the exterior frame of reference---the frame of an 
%observer who remains outside the horizon---the  process is  governed by 
%an S-matrix, connecting in-states to out-states. Then it is possible to 
%find a third operator in the Hilbert space of the outgoing degrees of 
%freedom (the Hawking radiation) which has the same statistics as
%$A(p)$. Call it $A_{**}({out})$.
Next, assume that in the exterior frame of reference---the frame of an 
observer who remains outside the horizon 
---the  process is  governed by an S-matrix, connecting in-states 
to out-states. Then it is possible to 
find an operator $A_{**}(in)$ in the Hilbert space for the in-states that has
the same statistics as a late time operator $A(q)$ which consists of 
the outgoing degrees of freedom (the Hawking radiation),  
\be
A_{**}({in}) = S^{\dag} A({in}) S =  S^{\dag}   U A_{*}(in) U^{\dag}    S.
\ee
%\be
%A_{**}({out}) = S^{\dag} A_*({in}) S =  S^{\dag}   U A(p) U^{\dag}    S.
%\ee
%
%In other words there is a mapping from the Hilbert space behind the 
%horizon to the Hilbert space of the outgoing degrees of freedom that 
%describes the Hawking radiation. The mapping is simply conjugation 
%by  $S^{\dag}   U $.  For every observable behind the horizon there is 
%a complementary observable composed out of the final evaporation
%products that has the same statistics. Of course the mapping is not an 
%easy one to decode. Black holes are very efficient ``scramblers" of 
%information \cite{scramble}.

In other words there is a mapping from an operator behind the 
horizon to an operator composed of the outgoing degrees 
of freedom.
%that describes the Hawking radiation. 
The mapping is simply conjugation by  $S^{\dag} U $.  
%For every observable behind the horizon there is 
%a complementary observable composed out of the final evaporation
%products that has the same statistics. 
Of course the mapping is not an easy one to decode. Black holes are 
very efficient ``scramblers" of information \cite{scramble}.

Consider a laboratory falling into a black hole. How much information can that laboratory contain? The answer is obvious: if every operator in the laboratory can also be represented in the Hilbert space of the outgoing evaporation products then the information in the laboratory cannot exceed the entropy of the black hole \cite{gavin}.
Let us define a concept of \ic \ for a system or subsystem\footnote{Not to be confused with channel capacity.}. The \ic \   is the maximum amount of information that the system can contain. Equivalently it is the maximum entropy of the system. For a quantum system it is the logarithm of the dimensionality of the Hilbert space of states needed to describe the system. What we have argued in the previous paragraph is that the \ic \ of any subsystem behind the horizon is bounded by the \ic \ of the black hole, i.e., the Bekenstein-Hawking entropy.

 The Census Taker's causal patch is bounded by a horizon. Part of that horizon is the hat itself but it also extends into the bulk geometry as in Figure 7. It should be clear that the Census Taker's region is the analog of the exterior of a black hole, and the portion of the multiverse beyond the horizon is the analog of the interior of the black hole.
 \begin{figure}
\begin{center}
\includegraphics[width=12cm]{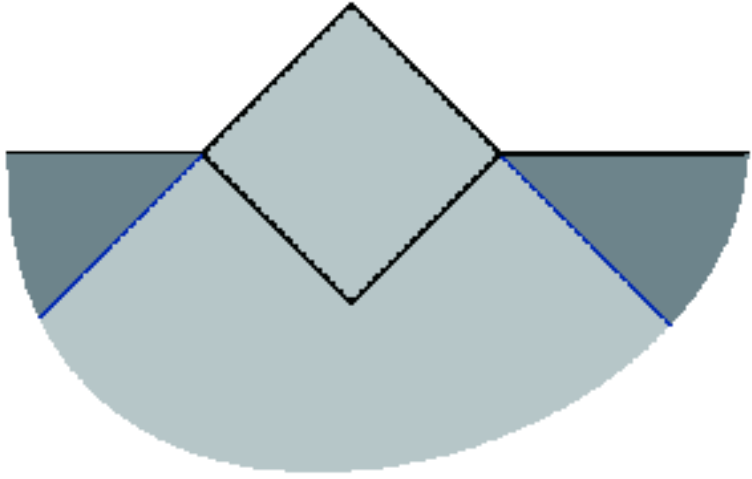}
\caption{The horizon of the CT consists of the hat and its continuation into the bulk} \label{hor}
\end{center}
\end{figure}
Most of the multiverse is behind the horizon and naively cannot be directly detected by the CT. But there is radiation coming into the CT's patch that is analogous to Hawking radiation.

The situation is similar to that of information behind the horizon of a black hole and its  evaporation products. In the practical sense the information becomes so scrambled that it is lost to an outside observer. In what follows we will conjecture a version of observer complementarity that applies to the CT's horizon.

We take the following assertions as given:

\bigskip
\it The degrees of freedom accessible to the CT are a complete description of the multiverse.

\bigskip
The Hilbert space of the CT's patch, i.e., the FRW region under the hat, is isomorphic to the Hilbert space of the multiverse. Operators outside the CT's horizon are complementary to operators within the horizon in the sense that they have the same statistics.\rm
\bigskip

At first sight these assumptions appear to be impossible since the FRW patch of the CT is a proper subset of the multiverse. It seems clear that the CT can never have enough information to decode the multiverse.

However, here is where the current setup differs from a black hole. The black hole can only store a finite amount of information---the entropy of the black hole---and an observer on the outside cannot collect more information than that. But in the case of the Census Taker, if he waits long enough there is no bound to the amount of information  that he can collect. In other words the \ic \ in the CT's patch is infinite. (It is worth noting that the \ic \ of a de Sitter space if it were stable would not be infinite.)

 The ``bulk" theory of the multiverse also requires  an unbounded  number of degrees of freedom to describe it. The phenomenon of eternal inflation will eventually populate the multiverse with an infinite number of events of every kind.
For these reasons,
counting and comparing the \ic  \ of the CT's patch and the \ic \ of the multiverse can only make sense for regularized versions of the theories.

There is one more point that is worth mentioning. Consider Figure 8.   \begin{figure}
\begin{center}
\includegraphics[width=12cm]{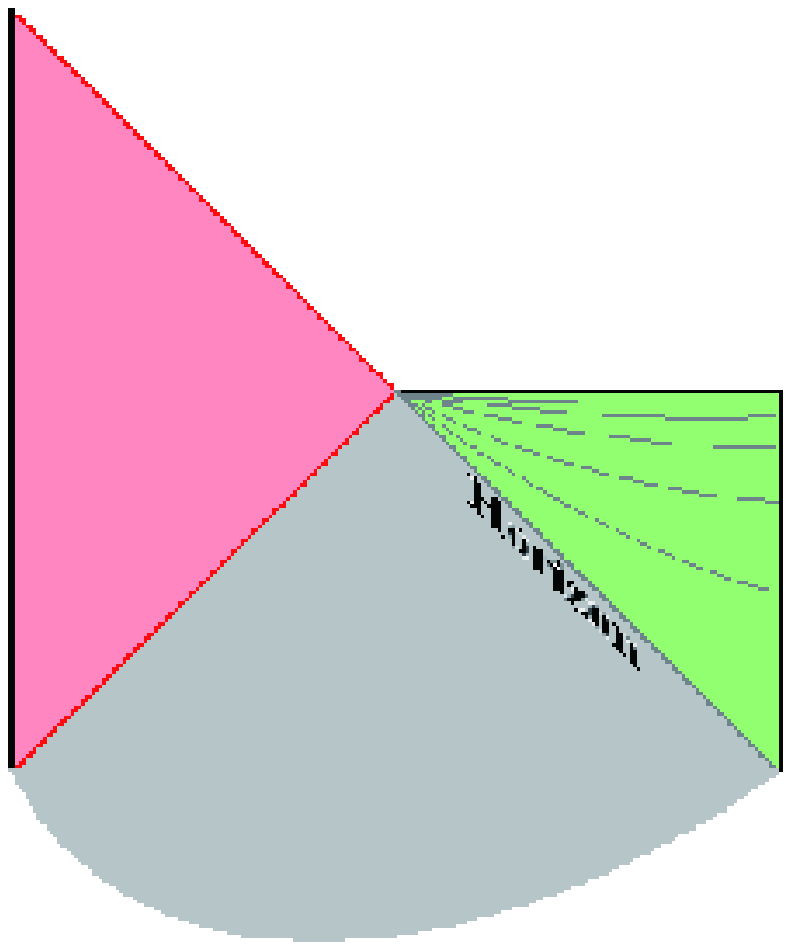}
\caption{The FRW patch and the portion of the Ancestor behind the horizon
share a common boundary $\Sigma.$} \label{horizons}
\end{center}
\end{figure}
The red region is the  FRW patch and the green region is the portion of the Ancestor vacuum which lies beyond the CT's horizon. Note that the two regions share a common boundary, namely
$\Sigma$. This suggests that the boundary-holographic theory describing  the Census Taker's patch may also be the  holographic description of the rest of the multiverse. 

The rest of this paper is about the duality of the boundary CFT and the FRW patch. It is independent of the conjectured observer complementarity relating it to the global description of the multiverse.

\setcounter{equation}{0}
\section{The Holographic Wheeler DeWitt Equation and the FSSY Conjecture}

The traditional approach to quantum cosmology---the \wdw \
equation---is the opposite of string theory: it is
 very flexible from the point of
view of background dependence---it doesn't require an an asymptotically
cold boundary condition, it can be formulated for a closed universe, a
flat or open FRW universe,  de Sitter space, or for that matter,
flat and \ads \ spacetime---but it is not
 a consistent quantum theory of gravity. It is based on the existence 
 of local ``bulk" degrees of freedom
 and therefore fails to
address the problems that String Theory and the  Holographic
Principle were designed to solve, namely, the huge over-counting of
degrees of freedom implicit in a local field theory.

Freivogel, Sekino, Susskind, and Yeh (FSSY)\cite{yeh} suggested a way out
of the dilemma: synthesize the \wdw \
philosophy with the Holographic Principle to construct a
Holographic \wdw \ theory. We will begin with a review of the
basics of conventional WDW; For a more complete treatment,
especially of infinite cosmologies, see \cite{banks et al}.

The ten equations of General Relativity take the form \be {\delta
\over \delta g_{\mu \nu}  } I =0 \label{GR equations} \ee where
$I$ is the Einstein action for gravity coupled to matter. The
canonical formulation of General Relativity makes use of a
time-space split \cite{ADM}. The six space-space components are
more or less conventional equations of motion, but the four
equations involving the time index have the form of constraints.
These four equations are written,
 \be
 H^{\mu}(x) =0.
 \label{adm}
 \ee
 They involve the space-space components of the metric $g_{nm}$, the matter fields
$\Phi$, and their conjugate momenta. The time component $H^0(x)$,
is a local Hamiltonian
 which ``pushes time forward" at the spatial point $x$. More generally, if  integrated
with a test function,
 \be
 \int d^3x \ f(x) \ H^0(x)
 \label{local push generator}
 \ee
it generates infinitesimal transformations of the form \be t \to t
\ + \ f(x).
 \label{local push}
\ee

Under certain conditions  $H^{0}$ can be integrated over space in
order to give a global Hamiltonian description. Since $H^{0}$
involves second space derivatives of $g_{nm}$, it is necessary to
integrate by parts in order  to bring the Hamiltonian to the
conventional form containing only first derivatives. In that case
the ADM equations can be written as \be \int d^3x \ H = E.
\label{H equals E} \ee The Hamiltonian density $H$ has a
conventional structure, quadratic in canonical momenta, and the
energy $E$ is given by a Gaussian surface integral over  spatial
infinity. The conditions which allow us to go from (\ref{adm})  to
(\ref{H equals E}) are satisfied in asymptotically cold
flat-space-time, as well as in \ads; in both cases global
Hamiltonian formulations exist. Indeed, in \ads \ the Hamiltonian
of the Holographic boundary description is identified with the ADM
energy, but, as we noted, cosmology, at least in its usual forms,
is never asymptotically cold. The only recourse for a canonical
description, is the local form of the equations (\ref{adm}).

When we pass from classical gravity to its quantum counterpart,
the usual generalization of the canonical equations (\ref{adm})
become the Wheeler DeWitt equations, \be H^{\mu} |\Psi \> = 0
\label{ H^mu Psi =0} \ee where the state vector $|\Psi \>$ is
represented by a wave functional that depends only on  the space
components of the metric $g_{mn}$, and the matter fields $\Phi$.

The first three equations \be H^{m} |\Psi \> = 0 \ \ \ (m = 1, \
2, \ 3) \label{momentum constraits} \ee have the interpretation
that the wave function is invariant under spatial diffeomorphisms,
 \be
 x^n \to x^n + f^n(x^m).
 \label{space diffeos}
 \ee
 In other words $\Psi(g_{mn}, \ \Phi)$ is a function of spatial invariants. These
equations are usually deemed to be the easy \wdw \ equations.

The difficult equation is the time component \be H^{0} |\Psi \> =
0. \label{hamiltonian constraint} \ee It represents invariance
under local, spatially varying, time translations. Not only is
equation (\ref{hamiltonian constraint}) difficult to solve; it is
difficult to  even formulate: the expression for $H^0$ is riddled
with factor ordering ambiguities. Nevertheless, as long as the
equations are not pushed into extreme quantum environments, they
can be useful.

\subsection{Wheeler DeWitt and the Emergence of Time}

Asymptotically cold backgrounds come equipped with a global
concept of time. But in the more interesting asymptotically warm
case, time is an approximate derived concept \cite{banks et
al,banks}, which emerges from the solutions to the \wdw \
equation. The perturbative method for solving (\ref{hamiltonian
constraint}) that was outlined in \cite{banks et al}, can be
adapted to the case of negative spatial curvature. We begin by
decomposing the spatial metric  into a constant curvature
background, and fluctuations. Since we will focus on open FRW
cosmology, the spatial curvature is negative, the space metric
having the form,
 \be ds^2 = a^2 \left(dR^2 + \sinh^2 R \ \ (d\theta^2 +
\sin^2 \theta d\phi^2) \right) + a^2 h_{mn}dx^m dx^n.
\label{complete spatial frw metric} \ee In (\ref{complete spatial
frw metric})\ $a$ is the usual FRW scale factor and the $x's$  are
$(R, \ \ \theta, \ \phi)$.

The first step in a semiclassical
expansion is the so-called mini-superspace approximation in which all fluctuations are ignored.  In lowest order, the Wheeler DeWitt wave function
depends only on the scale factor $a$. To carry out the leading
approximation in open FRW it is necessary to introduce an infrared
regulator which can be done by bounding the value of $R$, \be R <
R_0 \ \ (R_0 >> 1). \label{R=IR cutoff} \ee Let us also define the
total dimensionless coordinate-volume within the cutoff region, to
be $V_0$, \be V_0 = 4 \pi \int dR \sinh^2R \approx \12\pi
e^{2R_0}. \label{IR cutoff volume} \ee

The first (mini-superspace) approximation is described by the
action, \be L = {-a V_0\dot{a}^2 - V_0a   \over 2}. \label{miniL}
\ee Defining  $P$ to be the momentum conjugate to the scale factor
$a$, \be P = -a V_0 \dot{a} \label{P} \ee the Hamiltonian  $H$
is given by\footnote{The factor ordering in the first term is
ambiguous. We have chosen the simplest Hermitian factor ordering.},
\be H = {1\over 2 V_0}P{1\over  a}P - \12 V_0a \label{miniH}. \ee

Finally, using $P = -i \partial_a$, the first approximation to the
\wdw \ equation becomes, \be -\partial_a {1\over a}\partial_a \Psi
- V_0^2a \Psi=0. \label{miniwdw} \ee The equation  has the two
solutions, \be \Psi = \exp({\pm i V_0 a^2/2}), \label{minipsi} \ee
 corresponding to expanding and contraction universes; 
to see which is which we use
(\ref{P}). The expanding solution, labeled $\Psi_0$ is \be \Psi_0
= \exp({ -i V_0 a^2/2}). \label{psiexpanding} \ee From now on we
will only consider this branch.

At first sight there is something peculiar about (\ref{psiexpanding}). 
Multiplying
$V_0$ by $a^2$ seems like an odd operation. $V_0 a^3$ is the
proper volume, but what is $V_0 a^2$? In flat space it has no invariant significance; but in hyperbolic space its geometric meaning is
simple: it is just the proper area
of the boundary at $R_0$. One sees from the metric (\ref{FRW metric
in terms of T}) that the  coordinate volume $V_0$, and the
coordinate area $A_0$, of the boundary at $R_0$, are
(asymptotically) equal to one another, to within a factor of $2$,
\be A_0 = 2 V_0. \label{AV} \ee Thus the expression in the
exponent\footnote{In this subsection we are setting $4\pi G/3$ equal to 1.
Including this factor, the wave function is $\Psi_0=
\exp(-i ({3\over 4\pi})({A_0\over 4G}))$.} 
in (\ref{psiexpanding}) is $- i A_0/4$,
 where $A_0$ is the proper area  of the boundary at $R_0$, 
\be \Psi_0 = \exp({ -iA_0/4}). \label{areaphase} \ee

This is very suggestive. It is the first indication of a
holographic version of the WDW theory for an open FRW universe.

To go beyond the mini-superspace approximation one writes the wave
function as a product of $\Psi_0$, and a second factor $\psi(a, h,
\Phi)$ that depends on the fluctuations, \be \Psi(a, h, \Phi)=
\Psi_0 \ \psi(a, h, \Phi) = \exp({ -i A_0 a^2/4}) \ \psi(a, h,
\Phi). \label{fullwave} \ee By integrating the \wdw \ equation
over space, and substituting (\ref{fullwave}), an equation for
$\psi$ can be obtained, \be i\partial_a \psi - {1\over
A_0}\partial_a {1\over a}  \partial_a \psi = H_m \psi.
\label{smallwdw} \ee In this equation $H_m$ has the form of a
conventional Hamiltonian (quadratic in the momenta) for both
matter and metric fluctuations.

In the limit of large scale factor the term ${1\over
A_0}\partial_a  {1\over a}
\partial_a \psi$ becomes negligible and (\ref{smallwdw}) takes the form of a Schrodinger
equation, \be i\partial_a \psi  = H_m \psi. \label{schrod} \ee
Evidently the role of $a$ is not as a conventional observable, but
a parameter representing the unfolding of cosmic time. One does
not calculate its probability, but instead constrains it---perhaps
with a delta function or a Lagrange multiplier. As Banks has
emphasized \cite{banks}, in this limit, and maybe
only in this limit, the wave function $\psi$ has a conventional
interpretation as a probability amplitude.

As we have described it, the \wdw \ theory is a throwback to an
older view of quantum gravity based on the existence of bulk,
space-filling degrees of freedom. It has become clear that this is
a serious  overestimate of the capacity of space to contain
quantum information. The correct (holographic) counting of degrees
of freedom is in terms of the area of the boundary of space
\cite{hologram}. The question addressed by FSSY \cite{yeh} is how to combine the flexibility of the \wdw \ theory with the requirements of the Holographic Principle.

\subsection{The FSSY Conjecture}

The conjecture of \cite{yeh} \ is as follows:

\bigskip

\it The  holographic description of the FRW region under a hat is a Wheeler DeWitt theory in which the ordinary bulk degrees of freedom are replaced by degrees of freedom that reside on the asymptotic boundary of space i.e., on \sig. The Wheeler DeWitt wave function is a functional of those boundary degrees of freedom. Semiclassical bulk
degrees of freedom are approximate concepts reconstructed from the precise boundary
quantities.

\bigskip

Unlike the ADS case in which the boundary is asymptotically cold, the holographic degrees of freedom include a dynamical boundary metric.

\bigskip

As in the ADS/CFT correspondence \cite{susskind and witten},
it is useful to define a regulated boundary, $\Sigma_0$, at
$R=R_0$. In the regulated theory the number of degrees of freedom is proportional to the area of the regulated boundary in Planck units.

\rm

\bigskip

 In principle $R_0$ can depend on angular location on
$\Omega_2$. 
$$
R_0 = R_0(\Omega_2).
$$
In fact later we will discuss invariance under gauge
transformations of the form \be R\to R + f(\Omega_2). \label{R
shifts} \ee (The notation $ f(\Omega_2)$ indicating that $f$ is also a
function of location on $\Sigma_0$.)

Let us now consider the boundary degrees of freedom. In ADS/CFT the boundary theory is
typically a gauge theory and does not include gravitational degrees of freedom.
Asymptotic coldness  is the statement that the bulk fields are frozen at the boundary
and do not fluctuate. But in an asymptotically warm geometry  the boundary
geometry will fluctuate. For this reason the boundary degrees of freedom must include a
two dimensional spatial metric on $\Sigma_0$. The
induced spatial geometry of the boundary can always be described
in the conformal gauge in terms of  a Liouville  field
$U(\Omega_2),$ \be ds^2
=e^{2U(\Omega_2)}e^{2R_0(\Omega_2)}d\Omega_2^2. \label{liouville}
\ee

The Liouville field $U$ may be decomposed into  a homogeneous term $U_0$, and a
fluctuation. Obviously the homogeneous term can be identified with
the FRW scale factor,
 \be
 e^{U_0} = a.
 \label{homogeneous part
of U}
\ee
In section 6 we will give a more detailed definition of
the Liouville degree of freedom.

We also postulate  a collection of boundary ``matter" fields. The boundary matter
fields, $y$, are not the limits of the usual bulk fields $\Phi$,
but are analogous to the boundary gauge fields in the ADS/CFT
correspondence. In this paper we will not speculate on the
detailed form of  these boundary matter fields.

\subsection{The Wave Function}
In addition to  $U$ and $ y$, we assume a  local Hamiltonian
$H(x^i)$ that depends only on the boundary degrees of freedom (the
notation $x^i$ refers to coordinates of the boundary \sig), and a
wave function $\Psi(U, y)$, \be \Psi(U,y)= e^{-\12 S + iW}.
\label{psi equals exponential} \ee
 At every point of \sig, $\Psi$
satisfies \be H(x^i) \ \Psi(U,y) =0. \label{twoD wdw} \ee

 In equation (\ref{psi equals exponential}), $S(U,y)$ and $W(U,y)$ are real functionals
of the boundary fields. For reasons that will become clear, we
will call $S$ the action. However, $S$ should not in any way be
confused  with the four-dimensional Einstein action.

% The local Hamiltonian $H(x^i)$, and the imaginary term $W$ in 
%the exponent, play
% roles in determining the expectation values of canonical
%momenta, as well as the relation between scale factor and ordinary
%time. In this paper $H$ and $W$ will play secondary roles.

We make the following three preliminary assumptions about $S$ and $W$:

\begin{itemize}
\bigskip

\item    Both $S$ and $W$ are invariant under conformal
transformations of $\Sigma$. This follows from the $O(3,1)$ symmetry of the
background geometry.

\item The leading (non-derivative) term in the regulated form of
$W$ is $-A/4$ where $A$ is the proper area of $\Sigma_0$,
    \be
    W = -{1\over 4} \int_{\Sigma} e^{2R_0} e^{2U} +...
    \ee
    This follows from (\ref{areaphase}).
%%%
This term plays a role in determining the expectation values
of momentum conjugate to $U$, but will play secondary roles
in this paper. 

\item  $S$ has the form of  a local
 two dimensional Euclidean action on $\Sigma$. In other words it is an integral over
$\Sigma$, of densities that involve $U$, $y$, and their
derivatives with respect to $x^i$.

\end{itemize}
\bigskip

The first of these conditions is just a restatement of the
symmetry of the \cdl \ instanton. Later we will see that this
symmetry is  broken by a number of effects, including
the extremely interesting ``Persistence of Memory" discovered
by Garriga, Guth and Vilenkin \cite{garriga}.

The second condition follows from the bulk analysis described
earlier in equation (\ref{areaphase}). It allows us to make an
educated guess about the dependence of the local Hamiltonian
$H(x^i)$ on $U$. A simple form that reproduces (\ref{areaphase})
is \be H(x) = \12 e^{-2U} \pi_U^2 - {e^{4R_0}\over 8} e^{2U} + \ ...
\label{boundary H} \ee where $\pi_U$ is the momentum conjugate to
$U$. It is easily seen that the solution to the equation $H \Psi
=0$ has the form (\ref{areaphase}).

The highly nontrivial assumption is the third item---the locality
of the  action.  That the action $S$ is local  is far from
obvious; as a rule quantum field theory wave functions are
not local in this sense. In our opinion item three is the strongest
of our assumptions and the one most in need of confirmation. At
present our best evidence for the locality is the discrete tower
of boundary correlators, including a two-dimensional transverse, traceless, dimension-2 tensor-correlation function, discovered in \cite{yeh}. This is consistent with a dimension-2 energy momentum tensor---a necessary condition for a local field theory.

In
principle, much more information can be obtained from bulk
multi-point functions, continued to \sig. For example, correlation
functions of $h_{ij} $ would allow us to study the operator
product expansion of the energy-momentum tensor.

The assumption that $S$ is local is a very strong one,
but we mean it in a rather weak sense. One of the main points of
this paper is that there is a natural RG flow in cosmology (see
section 6). By locality we mean only that $S$ is in the basin of
attraction of a local field theory. If it is true, locality would
imply that the  measure \be \Psi^* \Psi = e^{-S} \label{psistar
psi = exp-S} \ee has the form of a  local two dimensional
Euclidean  field theory  with action $S$, and that the Census
Taker's observations could be organized not only by conformal
invariance but by conformal field theory.

%%%%%%%%%%%%%%%%%%%%%%%%%%%%%
%%%%%%%%%%%%%%%%%%%%%%%%%%%%%

 \setcounter{equation}{0}
  \section{Data}

  The conjectured locality of the action $S$ is based on data calculated by FSSY. The
background geometry studied in \cite{yeh} was the Minkowski
continuation of a thin-wall \cdl \ instanton, describing
transitions from the Ancestor vacuum to a  hatted supersymmetric
vacuum with zero cosmological constant.
In this section the data of \cite{yeh} will be reviewed.

We begin with some facts about three-dimensional
hyperbolic space and the solutions of its massless Laplace
equation. An important distinction is between normalizable modes
(NM) and non-normalizable modes (NNM). A minimally coupled scalar
field $\chi$ is sufficient to illustrate the important points.

The norm in hyperbolic space is defined in the obvious way: \be \<
\chi| \chi \> = \int dR d\Omega_2 \ \chi^2 \ \sinh^2 R
\label{norm} \ee In flat space, fields that tend to a constant at
infinity are on the edge on normalizability. With the help of the
delta function,  the concept of normalizability can be generalized
to continuum-normalizability, and  the constant ``zero mode" is
included in the spectrum of the wave operator, but in hyperbolic
space the normalization integral (\ref{norm})  is exponentially
divergent for constant $\chi$. The condition for normalizability
is that $\chi \to 0$ at least as fast as $e^{-R}$. The constant
mode is therefore non-normalizable.

Normalizable and non-normalizable modes have very different roles in the conventional
ADS/CFT correspondence. Normalizable modes are dynamical excitations with finite energy and can be produced
by events internal to the \ads. By contrast non-normalizable modes cannot be excited dynamically. Shifting the value of a
non-normalizable modes is equivalent to changing the boundary conditions from the bulk point of view, or changing
the Lagrangian from the boundary perspective. But, as we will see, in the
cosmological framework of FSSY,  asymptotic
warmness blurs this distinction.

The massless scalar correlator in the \cdl\ background stays finite
when the points approach the boundary~\cite{yeh}. In other words, 
non-normalizable modes (which have arbitrary angular dependence, and 
stay finite near the boundary) are excited. We will call this geometry 
asymptotic warm, in the sense that we cannot turn off perturbations at 
spatial infinity. 

The reason for this asymptotic warmness is the 
fact that the Euclidean version of the \cdl \ instanton is compact
(even though the spatial slice of the Lorentzian geometry is non-compact).
 The Euclidean metric is of the form,
\begin{equation}
ds^2=a^2(X)(dX^2+d\theta^2 +\sin^2\theta \Omega_2^2).
\label{EuclideanCDL} 
\end{equation}
Note that the Euclidean version of de Sitter space is a 4-dimensional
sphere (for which $a(X)=H^{-1}/\cosh X$, with $-\infty\le X\le \infty$, 
where $H$ is the Hubble 
constant of the ancestor vacuum). 
The Euclidean \cdl \ geometry in the thin-wall limit is a sphere cut at certain
value of $X$ and patched with a flat disc (for which $a(X)\propto 
H^{-1}e^{X}$) on the negative $X$ side. 
The open FRW universe (\ref{FRW metric in terms
of T}) is given by the analytic continuation 
\begin{equation}
X\to T+{\pi i/2}, \quad \theta\to iR,
\label{ac}
\end{equation}
from (\ref{EuclideanCDL}). (Note $e^{X}\to ie^{T}$,  
and that $S^3$ goes to  ${\cal H}_3$). 

%In the following, we review the calculation of correlators by
%analytic continuation from the Euclidean space.

Correlators are calculated in the Euclidean background 
(\ref{EuclideanCDL}) and analytically continued to the FRW universe, 
as we will review below.

\subsection{Scalars}

Correlation functions of massless (minimally coupled) scalars,
$\chi$, depend on time and on the dimensionless
geodesic distance between points on ${\cal{H}}_3$. In the limit in
which the points tend to the holographic boundary \sig \ at $R \to
\infty$, the geodesic distance between points $1$ and $2$ is given
by,
\be
l_{1,2} = R_1 + R_2 + \log{(1-\cos \alpha)}
\label{distance on
H}
\ee
where $\alpha$ is the angular distance on $S^2$
between $1$ and $2$. It follows on $O(3,1)$ symmetry grounds  that
the correlation function $\< \chi(1) \chi(2) \>$ has the form,
\bea
\< \chi(1) \chi(2) \> &=& G\left\{l_{1,2}, \ T_1, \ T_2, \right\} \cr &=&
G\left\{ R_1 + R_2 + \log{(1-\cos \alpha), \  T_1, \ T_2  }
\right\}.
 \label{scalar correlation in terms of T, alpha}
 \eea

Before discussing the data on the \cdl \ background, let us
consider the form of correlation functions for scalar fields in
\ads. We work in units in which the radius of the \ads \ is $1$.
By symmetry, the correlation function can only depend on $l$, the
proper distance between points. The
two-point function has the form
   \be
   \< \chi(1) \chi(2) \> \sim {e^{-(\Delta -1)l} \over \sinh l   }.
   \label{scalar correlators in ads}
   \ee
    In anti de Sitter space the dimension $\Delta$ is related to the mass of $\chi$ by
    \be
    \Delta (\Delta - 2) = m^2.
    \label{delta m relation}
    \ee

    We will be interested in the limit in which the two points  $1$ and $2$ approach the
boundary at $R \to \infty$. Using (\ref{distance on H}) gives
    \be
    \< \chi(1) \chi(2) \> \sim e^{-\Delta R_1} e^{-\Delta R_2}
    {(1 - \cos \alpha)}^{-\Delta}      .
   \label{asymptotic scalar correlators in ads}
   \ee

    It is well known that the ``infrared cutoff" $R$, in \ads, is equivalent to an
ultraviolet cutoff in the boundary Holographic description
\cite{susskind and witten}. The exponential factors, $\exp(-\Delta
R)$ in (\ref{asymptotic scalar correlators in ads}) have an important quantum field
theoretic meaning: they exactly correspond to
cutoff dependent wave function renormalization constants.
These factors are
normally stripped off when defining field theory correlators.
However, in this paper we will not remove them.

The remaining factor, $ {(1 - \cos \alpha)}^{-\Delta}$ is  the
conformally covariant correlation function of a boundary field of
dimension $\Delta$.

\begin{figure}
\begin{center}
\includegraphics[width=12cm]{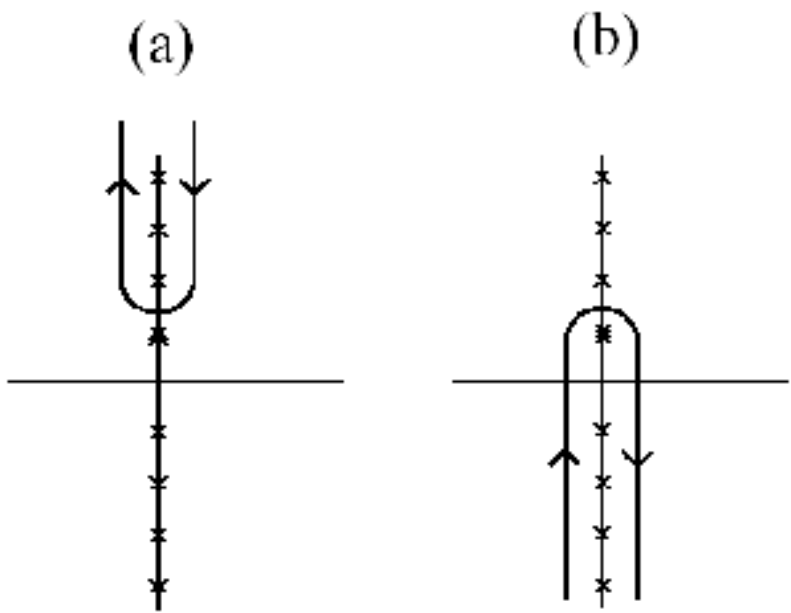}
\caption{Contours of integration for the two contributions $G_1, \
\ G_2$.} \label{6}
\end{center}
\end{figure}

Now let us briefly explain how to calculate the correlator on 
the \cdl\ background. 
Equation of motion for massless scalar $\chi$ in the
Euclidean \cdl\ geometry is
\begin{equation}
 \left[-\partial_X^2 +{a''(X)\over a(X)}-\nabla_S^2\right]
(a(X)\chi)=0,
\label{masslesseom}
\end{equation}
where $\nabla_S^2$ is the Laplacian on $S^3$. The Euclidean correlator 
is expressed as
\begin{equation}
 \langle \chi(X_1, 0) \chi (X_2,\theta)\rangle ={1\over a(X_1)a(X_2)}
\int_{C_1} {dk\over 2\pi} u_k^*(X_1)u_k(X_2)G_k(\theta),
\label{chichi}
\end{equation}
where $u_k(X)$ are the eigenfunctions of 
Schr\"{o}dinger operator: 
$[-\partial_X^2+a''(X)/a(X)]u_k(X)=(k^2+1)u_k(X)$,
and $G_k(\theta)=\sinh k(\pi-\theta)/(\sin \theta \sinh k\pi)$
is the Green's function on $S^3$ with mass $(k^2+1)$.
The correlator in the thin-wall limit can be written 
in terms of the reflection coefficient ${\cal R}(k)$ in
the potential $a''(X)/a(X)$.  
${\cal R}(k)$ has a pole at $k=i$, corresponding to a
bound state (we can see that $u_B(X)\propto a(X)$ 
is a bound state with eigenvalue $k=i$). 
%which corresponds to the bound state. 
The integration contour $C_1$ indicates that the integration
is done along the real axis and the contribution from 
the residue of the pole at $k=i$ is added (which is equivalent
to the integration along the contour (a)
in Figure~\ref{6})\footnote{We take this contour since 
we have to include the bound state mode in the complete 
set in the $X$ space. Otherwise, the correlator becomes singular at
$X\to-\infty$ (or $T\to-\infty$) even though the background
is smooth there.
The presence of a bound states in the Euclidean problem corresponds 
to the presence of non-normalizable modes in the FRW universe
as we will see below.}.
We can easily show that (\ref{chichi}) indeed satisfies (\ref{masslesseom})
with a delta function on the right hand side.

Performing the analytic continuation (\ref{ac}), we get
the correlator on the Lorentzian \cdl \ background,
\begin{equation}
 \langle \chi (T_1, 0)\chi (T_2, l)\rangle
= e^{-(T_1+T_2)}\int_{C_1} 
{dk\over 2\pi} \left( e^{ik(T_1-T_2+\pi)}+{\cal R}(k)e^{-ik(T_1+T_2)}
\right){\sin k l\over \sinh l\sinh k\pi}
\label{lorentziancorrelator} 
\end{equation}
where $l$ is the geodesic distance on ${\cal H}_3$. 

The first term, which has non-trivial dependence on ($T_1-T_2$), 
exists even when there is no bubble nucleation. Namely, the FRW 
region in the thin-wall limit
can be thought of as a part of flat Minkowski space (called
the Milne universe); the first term gives the correlator
in Minkowski space written in hyperbolic 
coordinates\footnote{We can do the integral for the first term 
in (\ref{lorentziancorrelator}) by contour deformation, and 
see that it agrees with the massless correlator in flat space,
$\langle\chi(1)\chi(2)\rangle=1/\{(\hat{T}_1-\hat{T}_2)^2
-|\hat{X}^a_1-\hat{X}^a_2|^2\}$, where $\hat{T}$, $\hat{X}$
are the Minkowski coordinates, related to the 
hyperbolic coordinates by $\hat{T}=e^{T}\cosh R$,
$\hat{X}^a=e^{T}\sinh R\Omega_a$.}. So we will call this
term the flat space piece. 
The second term which involves ${\cal R}(k)$ depend on
the details of bubble nucleation, such as the
Hubble constant of the false vacuum and the tension
of the domain wall.

In FSSY~\cite{yeh}, it was shown that the correlator 
on the \cdl\ background can be 
written as a sum of ADS correlators with definite 
masses (dimensions). By deforming the contour for the $k$ integration,
we get a discrete sum over residues at the poles. The integrand
has poles at integer multiples of $k=i$, and there is a 
double pole at $k=i$. In addition ${\cal R}(k)$ may have 
other singularities
in the lower half plane\footnote{Poles in the lower half plane
correspond to ``virtual states'' and resonances, which blow up
at $X\to \pm\infty$.}. 

Let us study the second term of (\ref{lorentziancorrelator}) which
involves ${\cal R}(k)$. In the region of our interest 
($l\to \infty$), the contour can be closed in the following way.
\be
e^{-(T_1 + T_2)}\left\{ \oint_a {dk \over 2\pi}
{\cal{R}}(k)  {e^{-ik(T_1 + T_2 -l)} \over  2 \ \sinh{l} \ \sinh{k \pi} }
+\oint_b {dk \over 2\pi}
{\cal{R}}(k)  {e^{-ik(T_1 + T_2 +l)} \over  2 \ \sinh{l} \ \sinh{k \pi} }
 \right\}
\label{contours}
\ee
where the contours $ \oint_a$ and $ \oint_b$ are shown in Figure~\ref{6}. 
FSSY ignored the 
contribution from poles in the lower half plane on the basis
that they are negligible at late time. In fact those terms
have significance that we will come back to, but first we will review
the terms studied in FSSY.

%The normalizable contribution, $G_1$, is  an infinite sum, each
%term having the form (\ref{asymptotic scalar correlators in ads})
%with $T$-dependent coefficients. For late times,

First, there is the normalizable contribution, $G_1$. This is 
an infinite sum, each term having the form (\ref{asymptotic 
scalar correlators in ads}) with $T$-dependent coefficients, 
%For late times,
\bea
   G_1 \eq \sum_{\Delta =2}^{\infty}G_{\Delta} e^{(\Delta -2)(T_1+T_2)} {e^{-(\Delta
-1)l} \over \sinh l} \cr
   %%%%%%%%%%
   %%%%%%%%%%%
  &\to& \sum_{\Delta =2}^{\infty}G_{\Delta} e^{(\Delta -2)(T_1+T_2)}  e^{-\Delta R_1}
e^{-\Delta R_2}
    {(1 - \cos \alpha)}^{-\Delta}
   \label{G1}
   \eea
   where $\Delta $ takes on integer values from $2$ to $\infty$, and $G_{\Delta}$ are a
series of constants which depend on the detailed CDL solution.

The connection with conformal field theory correlators is obvious;
equation (\ref{G1}) is a sum of correlation functions for fields
of definite dimension $\Delta$,  but with coefficients which
depend on the time $T$. (It should be emphasized that  the
dimensions $\Delta$  in the present context are not related to
bulk four dimensional masses.) 
%by (\ref{delta m relation}).) 
Note that the sum in (\ref{G1}) begins at $\Delta =2$, implying that
 every term falls at least as fast as $\exp{(-2R)}$
with respect to either argument. Thus every term is normalizable.

Let us now extrapolate (\ref{G1}) to the surface \sig. \sig \ can
be reached in two ways---the first being  to go out along  a
constant $T$ surface to $R = \infty$. Each term in the correlator
has a definite $R$ dependence which identifies its dimension.

  Another way to get to \sig \ is to first pass to light-like infinity, ${\cal{I}}^+$,
and then slide down the hat, along a light-like generator, until
reaching \sig. For this purpose it is useful to define light-cone
coordinates, $T^{\pm} = T \pm R$, \be G_1 = e^{-(T^+_1 +T^+_2)}
\sum_{\Delta} G_{\Delta} e^{(\Delta -1)(T^-_1 +T^-_2)}
(1-\cos \alpha)^{-\Delta}. \label{G1 in LC coords} \ee

We note that apart from the overall factor $ e^{-(T^+_1 +T^+_2)}$,
$G_1$ depends only on $T^-$, and therefore tends to a finite limit
on ${\cal{I}}^+$. If we strip that factor off, then the remaining
expression consists of a sum over CFT correlators, each
proportional to a fixed power of $e^{T^-}$. In the limit ($T^- \to
- \infty$) in which we pass to \sig, each term of fixed dimension
tends to zero as $e^{(\Delta -1)(T^-_1 +T^-_2)}$ with the
dimension-2 term dominating the others.

The second term in the scalar correlation function discussed by
FSSY consists of a single term, which comes from the double pole
at $k=i$, \be G_2 = {e^l \over \sinh{l}   }
(T_1 +T_2 +l) \to \left\{T^+_1 + T^+_2  + \log(1 - \cos
\alpha)\right \}.
 \label{G2}
\ee The contribution (\ref{G2}) does not have the form of a
correlator of a  conformal field of definite dimension.  To
understand its significance, consider a canonical massless scalar
field in two dimensions. On a two sphere the correlation function
is ultraviolet divergent and has the form \be \log\left\{\kappa^2
(1-\cos\alpha)\right\} \label{massless 2D correlator} \ee where
$\kappa$ is the ultraviolet regulator momentum. If the regulator
momentum varies with location on the sphere---for example in the
case of a lattice regulator with a variable lattice
spacing---formula (\ref{massless 2D correlator}) is replaced by \be
\log\left\{ (1-\cos\alpha)\right\} + \log \kappa_1 + \log \kappa_2.
\label{massless 2D correlator with variable cutoff} \ee Evidently
if we identify the UV cutoff $\kappa$ with $T^+ $,
 \be
\log \kappa = T^+, 
 \label{cutoff}
 \ee
 the expressions in (\ref{G2})  and (\ref{massless 2D correlator with variable cutoff})
are identical. The relation (\ref{cutoff}) is one of the central
themes of this paper, that as we will see, relates RG flow to the
observations of the Census Taker.

 That the UV cutoff of the 2D boundary theory depends on $R$ is very familiar from
the UV/IR connection \cite{susskind and witten} \ in \ads. In that
case the $T$ coordinate is absent and the log of the cutoff
momentum in the \cft \ would just be $R$. The additional time
dependent contribution in (\ref{cutoff}) will become clear later
when we discuss the Liouville field.

The logarithmic ultraviolet divergence in the correlator is a
signal that massless 2D scalars are ill defined; the well-defined
quantities being derivatives of the field\footnote{Exponentials of the field can also have well-defined dimensions. The dependence on the cutoff $\kappa$ in (\ref{cutoff}) becomes exponentiated. The resulting power law dependence on $\kappa$ is recognized as the wave function renormalization factor. }. When calculating
correlators of derivatives, the cutoff dependence disappears. Thus
for practical purposes, the only relevant term in (\ref{massless
2D correlator with variable cutoff}) is $\log{(1-\cos\alpha)}$.

The existence of a dimension-zero scalar field on \sig \ is a
surprise. It is obviously associated with bulk field-modes which
don't go zero for large $R$. Such modes are non-normalizable on
the hyperbolic plane, and are usually not included among the
dynamical variables in \ads.

In String Theory the only massless scalars in the supersymmetric  hatted vacua
would be moduli, which are expected to be ``fixed" in the
Ancestor. For that reason
 FSSY considered the effect of adding a four-dimensional mass term, $m^2 \chi^2$, in the
Ancestor vacuum. The mass term is assumed to vanish in the supersymmetric hatted vacuum.

The result on the boundary scalar was to shift
its dimension from $\Delta =0$ to $\Delta = \mu$ 
(when mass is small, $\mu\propto m^2$). If $m$ is sufficiently
small relative to the
Ancestor Hubble constant the corresponding mode stays non-normalizable.
However the
correlation function was not similar to those in $G_1$, each term
of  which had a  dependence on $T^-$. The dimension $\mu$ term depends
only on $T^+$:
 \be
 G_2 \to e^{-\mu T_1^+} e^{-\mu T_2^+} {(1 - \cos \alpha)^{-\mu}}. 
 \label{shifted G2}
 \ee

The two terms, (\ref{G1 in LC coords})  and (\ref{shifted G2})
depend on different combinations of the coordinates, $T^+$ and
$T^-$. It seems odd that there is one and only one term that
depends solely on $T^+$ and all the rest depend on $T^-$. In fact
the only reason for this was that FSSY ignored an entire tower of higher
dimension terms, which, like (\ref{shifted G2}), depend only on $T^+$. These terms all
come
 from the contour $\bf{b}$ in Figure~\ref{6}.

    From now on we will
group all  terms independent of $T^-$ into the single expression
$G_2$:
\be
G_2 = \sum_{\Delta'} \tilde{G}_{\Delta'} e^{-\Delta' (T^+_1
+T^+_2)}
(1-\cos \alpha)^{-\Delta'}. \label{G3}
\ee
The $\Delta'$ include
$\mu$, the positive integers, and contributions from 
whatever other poles in the lower half plane. 
%appear for $ik<1$. 
In the case $\mu = 0$, the leading term in $G_2$ is (\ref{G2}).

Finally there is the flat space piece coming from the
first term of (\ref{lorentziancorrelator}),
\begin{eqnarray}
 G_{\rm flat}&=& e^{-(T_1+T_2)}
\sum_{n=1}^{\infty}{e^{-n l}\over 2 \sinh l}
\left(e^{-n(T_1-T_2)} + e^{n(T_1-T_2)}\right)
\nonumber\\
&\rightarrow &\sum_{\Delta=2}^{\infty}\left(
e^{-\Delta T_{1}^{+}+(\Delta-2) T_{2}^{-}}
+e^{(\Delta-2) T_{1}^{-}-\Delta T_{2}^{+}}\right)
(1-\cos\alpha)^{-\Delta}.
\end{eqnarray}
Here the $T^{+}$ dependence at one point is combined with the 
$T^{-}$ dependence at the other point. 

We will return to the two terms $G_1$ and $G_2$  in section 6.5.

\subsection{Metric Fluctuations}

To prove that there is  a local field theory on \sig, the most
important test is the existence of an energy-momentum tensor. In
the ADS/CFT correspondence, the boundary energy-momentum tensor is
intimately related to the bulk metric fluctuations. We assume a
similar connection between bulk and boundary fields in the present
context. In FSSY, metrical fluctuations were studied in a
particular gauge which we will call the \it  Spatially
Transverse-Traceless \rm (STT) gauge. The coordinates of region I
can be divided into FRW time, $T$, and space $x^m$ where $m = 1, \
2, \ 3.$ The STT gauge for  metric fluctuations is defined by \bea
\nabla^m h_{mn} \eq 0, \cr h_m^m  \eq 0. \label{STT gauge} \eea In
these equations, the index is raised
with the aid of the background metric (\ref{FRW metric in terms of
T}). The main benefit of the STT gauge is that  metric
fluctuations satisfy minimally coupled massless scalar
equations, and the correlation functions are similar to $G_1$ and
$G_2$. However the index structure is rather involved. We define
the correlator, \bea \< h^{\mu}_{\nu} h^{\sigma}_{\tau} \> \eq G
\left\{^{\mu \sigma}_{\nu \tau}   \right\} \cr \eq G_1
\left\{^{\mu \sigma}_{\nu \tau} \right\} + G_2 \left\{^{\mu
\sigma}_{\nu \tau} \right\}. \label{gravity correlaror G1+G2} \eea

The complicated index structure of  $ G $ was worked out in detail
in FSSY. In this paper we quote only the results of interest---in
particular those involving  elements of $G \left\{^{\mu
\sigma}_{\nu \tau}   \right\}$ in which all indeces  lie in the
two-sphere $\Omega_2.$  
%Thus we consider the correlation function
%$G_1 \left\{^{i k}_{j l} \right\}$.
We consider the part $G_1 \left\{^{i k}_{j l} \right\}$, which
contains a term with dimension 2. This is the dimension that
energy-momentum tensor in a two-dimensional 
conformal field theory should have\footnote{$G_2$ does not contain $\Delta=2$ 
term because ${\cal R}(k)$ is zero at $k=-i$ which
would correspond to $\Delta=2$~\cite{yeh}.}.

As in the scalar case, $G_1 $ consists of an infinite sum of
correlators, each corresponding to a field of dimension $\Delta =
2,\ 3, \ 4,...$. The asymptotic $T$ and $R$  dependence of the
terms is identical to the scalar case, and the
first term has $\Delta = 2$. 
%This is particularly interesting because it is the dimension of 
%the energy-momentum tensor of a two-dimensional boundary \cft. 
Once again this term is also time-independent.

After isolating the dimension-two term and stripping off the
factors $\exp{(-2R)}$, the resulting correlator is called $G_1
\left\{^{i k}_{j l}   \right\}|_{\Delta =2}$. The calculations of
FSSY revealed that this term is two-dimensionally
 traceless, and transverse,  \bea G_1 \left\{^{i k}_{i l}
\right\}|_{\Delta =2} = G_1 \left\{^{i k}_{j k} \right\}|_{\Delta
=2} &=&0 \cr
%%%%%%%%%%%%%%%%%%%%%%%%%%%%%%%%%%%%%%
%%%%%%%%%%%%%%%%%%%%%%%%%%%%%%%%%%%%%%
\nabla_i G_1 \left\{^{i k}_{j l}   \right\}|_{\Delta =2} &=&0.
\label{2d transverse traceless} \eea

Equation (\ref{2d transverse traceless}) is the clue that, when
combined with the dimension-2 behavior of $ G_1\left\{^{i k}_{i l}
\right\}|_{\Delta =2}$, hints at a local theory on \sig. It
insures that it has the precise form of a two-point function for
an energy-momentum tensor in a conformal field theory. The only
ambiguity is the numerical coefficient connecting $G_1 \left\{^{i
k}_{j l} \right\}|_{\Delta =2}$ with $\<T^i_j T^k_l \>$. We will
return to this coefficient momentarily.

The existence of a transverse, traceless, dimension-two operator
is a necessary condition for the boundary theory on \sig \ to be
local: at the moment it is  our main evidence. But there is
certainly more that can be learned by computing multipoint
functions. For example, from the three-point function $\<hhh\>$ it
should be possible to verify the operator product expansion and the
Virasoro algebra for the energy-momentum tensor.

Dimensional analysis allows us to estimate the missing coefficient
connecting the metric fluctuations with $T^i_j$, and at the same
time determine the central charge $c$.  In \cite{yeh}  \ we found
$c$ to be of order the horizon entropy of the Ancestor vacuum. We
repeat the argument here:

Assume that the (bulk) metric fluctuation $h$ has canonical
normalization, i.e., it has bulk mass dimension $1$ and a
canonical kinetic term. Either dimensional analysis or explicit
calculation of the two point function $\<hh\>$ shows that it is
proportional to square of the Ancestor Hubble constant, \be \< h
h\> \sim H^2. \label{hh normalization} \ee Knowing that the  three
point function $\< hhh\>$ must contain  a factor of the
gravitational coupling (Planck Length) $l_p$, it can also be
estimated by dimensional analysis, \be \< h hh\> \sim l_p H^4.
\label{hhh normalization} \ee Now assume that the 2D
energy-momentum tensor is proportional to the boundary
dimension-two part of $h$, i.e., the part that varies like
$e^{-2R}$. Schematically, we write \be T = q h \label{T=qh} \ee with $q$
being a numerical constant. It follows that \bea \<TT\> &\sim& q^2
H^2 \cr \<TTT\> &\sim& q^3 l_p H^4. \label{TT and TTT correlators}
\eea

Lastly, we use the fact that the ratio of the two and three point
functions is parametrically independent of $l_p$ and $H$ because
it is controlled by the classical algebra of diffeomorphisms:
$[T,T] = T$. Putting these elements together we find, \be \<TT\>
\sim {1 \over l_p^2 H^2} \label{final TT} \ee Since we already
know that the correlation function has the correct form, including
the short distance singularity, we can assume that the right hand
side of (\ref{final TT}) also gives the central charge. It can be
written in the rather  suggestive form: \be c \sim Area  / G  \  \
\  \  (G = l_p^2) \label{central charge} \ee where $Area$ refers
to the horizon of the Ancestor vacuum.   In other words, the
central charge of the hypothetical CFT is proportional to the
 horizon entropy of the Ancestor.

\subsection{Dimension Zero Term}

The term $G_2 \left\{^{i k}_{j l}   \right\}$ begins with a term,
which  like its scalar counterpart, has a non-vanishing limit on
\sig. It is expressed in terms of a standard 2D bi-tensor
$t\left\{^{i k}_{j l}   \right\}$ which is traceless and
transverse in the two dimensional sense. If the correlation
function were  given just by $t\left\{^{i k}_{j l} \right\}$, it
would be a pure gauge artifact. One can see this by considering
the linearized expression for the 2D curvature-scalar  $C$, \be C
= \nabla_i \nabla_j h^{ij}-2\nabla^i \nabla_i \ Tr \ h.
\label{curvature} \ee The 2D curvature associated with a traceless
transverse fluctuation vanishes, and since $t\left\{^{i k}_{j l}
\right\}$ by itself is traceless-transverse with respect to both
points, it would be  pure gauge if it appeared by itself.

However, the actual correlation function $G_2 \left\{^{i k}_{j l}
\right\}$  is given by \be G_2 \left\{^{i k}_{j l}   \right\} =
t\left\{^{i k}_{j l}   \right\} \left\{R_1 + T_1 + R_2 + T_2 +
\log({1 - \cos{\alpha}})\right\} \label{G2(h)} \ee The linear
terms in $R+T$, being proportional to $t\left\{^{i k}_{j l}
\right\}$ are pure gauge, but the finite term \be t\left\{^{i
k}_{j l}   \right\}
 \log({1 - \cos{\alpha}})
\label{finite part of hh} \ee gives rise to a non-trivial 2D
curvature-curvature correlation function of the form \be \<CC\> =
(1-\cos{\alpha})^{-2} \label{curvature correlator}. \ee One
difference between   the metric fluctuation $h$, and the scalar
field $\chi$, is that we cannot add a mass term for $h$ in the
Ancestor vacuum to shift its dimension.

Finally,  as in the scalar case, there is a tower of higher
dimension terms in the tensor correlator, $G_2 \left\{^{i k}_{j l}
\right\}$ that only depend on $T^+$.

The existence of a zero dimensional term in $G_2 \left\{^{i k}_{j
l}   \right\}$, which remains finite in the limit $R\to \infty$
indicates that fluctuations in the boundary geometry---fluctuations
which are due to the asymptotic warmness---cannot be ignored. One
might expect that in some way these fluctuations are connected
with the field $U$ that we encountered in the Holographic version
of the \wdw \ equation. In the next section we will elaborate on
this connection.

\subsection{Three-point functions}
Three-point functions in the \cdl \ geometry can also be calculated 
by analytic continuation from the Euclidean space. 
They take the form of ADS three-point functions with mass
of the propagators summed over. Let us illustrate this by taking
the tree-level three-point function for a massless field $\chi$
with an interaction term $\int d^4x \sqrt{g}\chi^3$, as an example. 

The Euclidean three-point function is given by
integrating the vertex over the whole Euclidean space,
\begin{equation}
 \langle \chi (x_1)\chi (x_2)\chi (x_3)\rangle
=\int d^{4}x_0 \langle \chi (x_1)\chi (x_0)\rangle
\langle \chi (x_2)\chi (x_0)\rangle \langle \chi (x_3)\chi (x_0)\rangle,
\end{equation}
where the subscript 0 denotes the coordinates of the vertex.
The propagators $\langle \chi \chi\rangle$ are given by the 
two-point functions (\ref{chichi}). 
In the thin-wall background, $\langle \chi \chi \rangle$ is written
in terms of the reflection or the transmission coefficients ${\cal R}(k)$,
${\cal T}(k)$. The external points are put on the flat side, since
we want three-point functions in the flat FRW universe. 
When the vertex is on the flat side, 
\begin{equation}
 \langle \chi (X_1, \Omega_1)\chi (X_0, \Omega_0)\rangle
={1\over a(X_1)a(X_0)}\int_{C_1} {dk\over 2\pi}
(e^{ik(X_1-X_0)}+{\cal R}(k)e^{-ik(X_1+X_0)})G_k(\Omega_{10}),  
\end{equation}
and when the vertex is on the de Sitter side, 
\begin{equation}
 \langle \chi (X_1, \Omega_1)\chi (X_0, \Omega_0)\rangle
={1\over a(X_1)a(X_0)}\int_{C_1} {dk\over 2\pi}
{k+i\tanh X_0\over k+i}{\cal T}(k)e^{-ik(X_1-X_0)}G_k(\Omega_{10}),
\end{equation}
where $\Omega_{1}$, $\Omega_{0}$ denote the positions on $S^3$, and
$\Omega_{10}$ is the geodesic distance on $S^3$. 

Performing the integration over the Euclidean time ($X_0$) 
of the vertex, we get
\begin{eqnarray}
&&\hspace{-1cm}\langle \chi (X_1,\Omega_1)\chi (X_2, \Omega_2)
\chi (X_3, \Omega_3)\rangle={1\over a(X_1)a(X_2)a(X_3)}
\int_{C_1} {dk_1\over 2\pi}\int_{C_1} {dk_2\over 2\pi}\int_{C_1} 
{dk_3\over 2\pi}G_{k_1k_2k_3}(\Omega_1, \Omega_2, \Omega_3)\nonumber\\
&&\cdot \left[(e^{ik_1X_1}+{\cal R}(k_1)e^{-k_1X_1})
\cdot(2)\cdot(3)\cdot c^{(\rm flat)}_{k_1k_2k_3}
+{\cal T}(k_1)e^{-ik_1X_1}\cdot(2)\cdot(3)\cdot c^{(\rm dS)}_{k_1k_2k_3}\right]
\label{Euclidean3pt}
\end{eqnarray}
where $(2)$ and $(3)$ denote the factors given by replacing the 
subscripts in the previous factors by 2 and 3. 
$c^{(\rm flat)}_{k_1k_2k_3}$ and $c^{(\rm dS)}_{k_1k_2k_3}$
are the coefficients we get when the vertex on the flat side
and de Sitter side, respectively. 
$G_{k_1k_2k_3}(\Omega_1, \Omega_2, \Omega_3)$ is a three-point 
function for massive fields on $S^3$, 
\begin{equation}
 G_{k_1k_2k_3}(\Omega_1, \Omega_2, \Omega_3)
=\int d^{3}\Omega_0G_{k_1}(\Omega_{10})
G_{k_2}(\Omega_{20})G_{k_3}(\Omega_{30}).
\end{equation}

To get the three-point function in the FRW universe from 
(\ref{Euclidean3pt}), we analytically continue the coordinates
of the external
points by (\ref{ac}), do the $k$ integrals by closing the 
contour in the appropriate directions, and rotate the integration
contour for $\int d^3\Omega_0$ to get an integral over ${\cal H}_3$. 
The final result is a sum of massive three-point functions
on ${\cal H}_3$ (3-dimensional Euclidean ADS).
The mass (dimension) for each ADS propagator is summed over;
as in the two-point function, the dimension takes integer values 
starting from 2, and other possible values coming from the poles 
of ${\cal R}(k)$ or ${\cal T}(k)$.

The detailed structure of three-point functions is 
under study. Three-point functions involving a graviton
will tell us the precise value of the central charge,
as mentioned in Section 5.2. Also, the study of operator
algebra will provide non-trivial consistency checks for
our proposal of identifying a bulk field with a tower of
CFT operators.
%They will also provide non-trivial tests for the
%identification of a bulk field with a tower of the CFT operators
%that we have been assuming. 
\bigskip

That is the data about correlation functions on the boundary
sphere \sig \ that form the basis for our conjecture that there
exists a local holographic boundary description of the open FRW
universe. There are a number of related puzzles that  this data
raises: First, how does time emerge from a Euclidean QFT? The bulk
coordinate $R$ can be identified with scale size just as in
ADS/CFT\footnote{This is often described by saying that $R$ is related to the renormalization group flow parameter. } but the origin of time requires a new mechanism.

The second puzzle concerns the number of degrees of freedom in the
boundary theory. The fact that the central charge is the entropy
of the Ancestor suggests that there are only enough degrees of
freedom to describe the false vacuum and not the much large number
needed for the open FRW universe at late time. The resolution of both puzzles
involves the Liouville field.

%%%%%%%%%%%%%%%%%%%%%%%%%%%%%%%%
%%%%%%%%%%%%%%%%%%%%%%%%%%%%%%%%

\setcounter{equation}{0}
\section{Liouville Theory}

\subsection{Breaking Free of the STT Gauge}

The existence of a Liouville sector describing metrical
fluctuations on \sig \ seems dictated by both the Holographic \wdw
\ theory and from the data of the previous section. It is clear
that the Liouville field is  somehow connected with the
non-normalizable metric fluctuations whose correlations are
contained in (\ref{G2(h)}), although the connection is somewhat
obscured by the choice of gauge in \cite{yeh}.  In the  STT gauge
the fluctuations $h$ are traceless, but not transverse (in the 2D
sense). From the viewpoint of 2D geometry they are not pure gauge
as can be seen from the fact that the 2D curvature correlation
does not vanish. One might be tempted to identify the Liouville
mode with the zero-dimension piece of (\ref{G2(h)}). To do so
would of course require a coordinate transformation on $\Omega_2$
in order to bring the fluctuation $h_i^j$ to the ``conformal" form
${\tilde{h}} \delta_i^j$.

This identification may be useful but it is not consistent with
the \wdw \ philosophy. The Liouville field $U$ that appears in the
\wdw \ wave function  is not tied to any specific spatial gauge.
Indeed, the wave function is required to be invariant under gauge
transformations, \be x^{\mu} \to x^{\mu} + f^{\mu}(x) \label{gauge
transform on x} \ee under which the metric transforms: \be g_{\mu
\nu} \to g_{\mu \nu} +\nabla_{\mu}f_{\nu} +\nabla_{\nu}f_{\mu}.
\label{gauge transform of g munu} \ee

%%%%%%%%%%%%%%%%%
Let's consider the effect of such transformations on the boundary
limit of $h_{ij}$. The components of $f$ along the directions in
\sig \ induce 2D coordinate transformation under which $h$
transforms conventionally. Invariance under these transformations
merely mean that the action $S$ must be a function of 2D
invariants.
%%%%%%%%%%%%%%%%%%%%%%%%%%%%%%%%%%%%%%%%%%%%%%%%%%%%%%%%%%%%%

Invariance under the shifts $f^R$ and $f^T$ are more interesting.
In particular the combination $f^+ = f^R + f^T$ generates
non-trivial transformations of the boundary metric $h_{ij} $. An
easy calculation shows that, \be h_i^j \to h_i^j +
f^+(\Omega_2)\delta_i^j. \label{trans of h under f+} \ee

In other words, shift transformations $f^{+}$, induce Weyl
re-scalings of the boundary metric. This prompts us to modify the
definition of  the Liouville field from \be U = T + \tilde{h}
\label{not liouville} \ee to \be U= T + \tilde{h} + f^+.
\label{Liouville = T + h + f} \ee

One might wonder about the meaning of  an equation such as
(\ref{Liouville = T + h + f}). The left side of of the equation is
supposed to be a dynamical field on \sig,  but the right side
contains an arbitrary function $f^+$. The point is that in the
Wheeler DeWitt formalism the wave function must be invariant under
shifts, but in the original analysis of FSSY  a specific gauge was
chosen. Thus, in order to render the wave function gauge
invariant, one must allow the shift $f^+ $ to be an integration
variable, giving it the status of a  dynamical field.

A similar example is familiar from ordinary gauge theories. The
analog of the \wdw \ gauge-free formalism would be the unfixed
theory in which one integrates over the time component of the
vector potential.  The analog of the STT gauge would be the
Coulomb gauge. To go from one to the other we would perform the
gauge transformation \be A_0 \to A_0 + \partial_0 \phi.
\label{gauge transform} \ee Integrating over the gauge function
$\phi$ in the path integral would restore the gauge invariance
that was given up by fixing Coulomb gauge.

Returning to the Liouville field, since both $\tilde{h}$ and $f$
are linearized fluctuation variables,  we see that the classical
part of $U$ is still the FRW conformal time.

One important point: because the effect of the shift $f^+$ is
restricted to the trace of $h$, it does not influence the
traceless-transverse (dimension-two) part of the metric
fluctuation, and the original identification of the 2D
energy-momentum tensor is unaffected.

Finally, invariance under the shift $f^-$ is trivial in this
order, at least for the thin wall geometry. The reason is that in
the background geometry, the area does not vary along the $T^-$
direction.

Given that the boundary theory is local, and includes a boundary
metric, it is constrained by the  rules of two-dimensional quantum
gravity laid down long ago by Polyakov \cite{polyakov}. Let us
review those rules for the case of a conformal ``matter" field
theory coupled to a Liouville field. Two-dimensional coordinate
invariance implies that the central charge of the Liouville sector
cancels the central charge of all other  fields. We have argued in
\cite{yeh} (and in section 5) that the central charge of the
matter sector is of order
 the horizon area of the Ancestor vacuum, measured in Planck Units. It is obvious from
the 4-dimensional bulk viewpoint that the semiclassical analysis
that we have relied on, only makes sense when  the Hubble radius
is much larger than the Planck scale. Thus we we take the central
charge of matter to satisfy $c>>1$. As a consequence, the central
charge of the Liouville sector, $c_L$, must be large and negative.
Unsurprisingly,  the negative value of $c$ is the origin of the
emergence of time.

The formal development of Liouville theory begins by defining two
metrics on $\Omega_2$. The first is what we will call the reference
metric ${\hat{g}}_{ij}$. Apart from an appropriate degree
smoothness, and the assumption of Euclidean signature, the
reference metric is arbitrary but fixed. In particular it is not
integrated over in the path integral. Moreover, physical
observables must be independent of ${\hat{g}}_{ij}$

The other metric is the ``real" metric denoted by $g_{ij}$. The
purpose of the reference metric is merely to implement a degree of
gauge fixing. Thus one assumes that the real metric has the form,
\be 
g_{ij} = e^{2U}{\hat{g}}_{ij}. 
\label{mets}
\ee 
The real metric---that is to
say $U$---is a dynamical variable to be integrated over.

 For positive $c_L$ the Liouville Lagrangian is
 \be
L_L= {Q^2 \sqrt{\hat{g}} \over 8\pi}\left\{ \hat{\nabla}  U
\hat{\nabla}  U  + \hat{R} U \right\} \label{L-lagrangian}
 \ee
 where $ \hat{R}, \ \ \hat{\nabla} $, all refer to the  sphere
$\Omega_2$, with  metric  ${\hat{g}}$. The constant $Q$ is related
to the central charge $c_L$ by
 \be
 Q^2 = {c_L  \over 3  }
 \label{Q}
 \ee
 The two dimensional cosmological constant has been set to zero for the moment, but it
will return to play a surprising role. For future reference we
note that the cosmological term, had we included it, would have
had the form,
 \be
 L_{cc} = \sqrt{\hat{g}}\lambda e^{2U}.
 \label{Lcc}
 \ee

It is useful to define a field $\phi = QU $ in order to bring the
kinetic term to canonical form. One finds,
 \be
L_L= {\sqrt{\hat{g}} \over 8\pi}\left\{ \hat{\nabla}  \phi
\hat{\nabla}  \phi + Q\hat{R} \phi \right\} \label{canon
L-lagrangian}
 \ee
 and, had we included a cosmological term, it  would be
  \be
 L_{cc} =\sqrt{\hat{g}} \lambda \exp{ 2\phi \over Q}.
 \label{canon Lcc}
 \ee

By comparison with the case of positive $c_L$, very little is
rigorously understood about Liouville theory with negative central
charge. In this paper we will make a  leap of faith: we assume that the theory can be
defined by analytic continuation from positive $c_L$. To that end
we note that the only place that the central charge enters
(\ref{canon L-lagrangian}) and (\ref{canon Lcc}) is through the
constants $Q$ and $\gamma$, both of which become imaginary when
$c_L$ becomes negative. Let us define, \be Q = i\cal{Q}.
\label{calQ} \ee

Equations (\ref{canon L-lagrangian}) and (\ref{canon Lcc}) become,
\bea L_L \eq {\sqrt{\hat{g}} \over 8\pi}\left\{ \hat{\nabla}  \phi
\hat{\nabla}  \phi + i{\cal{Q}}\hat{R} \phi \right\} \cr \eq
{\sqrt{\hat{g}} \over 8\pi}\left\{ \hat{\nabla}  \phi \hat{\nabla}
\phi + 2i{\cal{Q}} \phi \right\} \label{imaginary lagrangian}
 \eea
 (where we have used $\hat{R} =2$),
 and
  \be
 L_{cc} =\sqrt{\hat{g}} \lambda \exp{-2i \phi \over \cal{Q}}.
 %\label{canon Lcc}
 \ee

Let us come now to the role of $\lambda$. First of all   $\lambda
$ has nothing to do with the four-dimensional cosmological
constant, either in the FRW patch or the Ancestor vacuum.
Furthermore it is not a constant in the action of the boundary
theory. Its proper
  role is as a \it Lagrange multiplier \rm that serves to specify the time $T$,
or more exactly, the global scale factor. The procedure is
motivated by the \wdw \ procedure of identifying the scale factor
with time. In the present case of the thin-wall limit, we identify
$\exp{2U}$ with $\exp{2T}$. Thus we insert  a $\delta$ function in
the path integral, \be \delta \left(\int  {\sqrt{\hat{g}}} (e^{2U}
- e^{2T})\right) =
\int dz \exp{iz\left(\int  {\sqrt{\hat{g}}} (e^{2U} -
e^{2T})\right)} \label{delta fnct} \ee

The path integral (which now includes an integration over the
imaginary 2D cosmological constant $z$) involves the action \be
L_L + L_{cc}  =
 {\sqrt{\hat{g}} \over 8\pi}\left\{ \hat{\nabla} \phi \hat{\nabla}  \phi + 2i{\cal{Q}}
\phi  + 8 \pi i z \exp{-2i \phi \over \cal{Q}} -8\pi i z
e^{2T}\right\}
 \label{total L}
\ee

The action (\ref{total L}) has a saddle point\footnote{It should be noted that the saddle point only occurs for negative central charge.} when the potential
 \be
V  = 2i{\cal{Q}} \phi  + 8 \pi i z \exp{-2i \phi \over \cal{Q}}
-8\pi i z e^{2T}
 \label{V}
\ee
 is stationary; this occurs at,
\bea \exp{-2i \phi \over \cal{Q}} \eq e^{2T} \cr z \eq i
{{\cal{Q}}^2 \over 8\pi  } e^{-2T} \label{presaddle} \eea or in
terms of the original variables, \bea e^{2U} \eq e^{2T} \cr
\lambda \eq  {{\cal{Q}}^2 \over 8\pi  } e^{-2T} \label{saddle}
\eea

Once $\lambda$ has been determined in terms of $T$ by (\ref{saddle}), the
Liouville theory with that value of $\lambda$ determines
expectation values of the remaining variables as functions of the
time.

Thus, as we mentioned earlier, the cosmological constant is
not a constant of the theory but rather a parameter that we scan
in order to vary the cosmic time.

 \subsection{Liouville, Renormalization, and Correlation Functions}

 \subsubsection{Preliminaries}

  There are two preliminary discussions that will help us understand the application of
Liouville theory to cosmic holography. The first is about the
ADS/CFT connection between the bulk coordinate $R$, and
renormalization-group-running of the boundary field theory. There
are three important length scales in every quantum field theory.
The first is the ``low energy scale;" in the present case the low
energy scale is the radius of the sphere which we will call $L$.

   The second  is the ``bare" cutoff scale---where the underlying theory is
   prescribed.
Call the bare scale\footnote{The lattice spacing  $a$  should not
be confused with the FRW scale factor, also called $a$.} $a$.

The bare input is a collection of degrees of freedom,
and an action coupling them. In a lattice gauge theory the degrees
of freedom are site and link variables, and  the couplings are
nearest neighbor to insure locality\footnote{Nearest neighbor is
common but not absolutely essential. However this subtlety is  not
important for us.} In a ferromagnet they are spins situated on the
sites of a crystal lattice.

The previous  two  scales have  obvious physical meaning but the
third scale is arbitrary: a sliding scale called the
renormalization or reference scale. We  denote it by $\delta$. The
reference scale is assumed to be much smaller than $L$ and much
larger than $a$, but otherwise it is arbitrary.  It helps to keep
a concrete model in  mind. Instead of  a regular lattice,
introduce a ``dust" of points with average spacing $a$. It is  not
essential that $a$ be uniform on the sphere. Thus the spacing of
dust points
 is  a function of position, $a(\Omega_2)$.  The degrees of
freedom on the dust-points, and their nearest-neighbor couplings,
will  be left implicit.

Next we introduce a second dust at larger spacing, $\delta$. The
$\delta$-dust provides the reference scale. It is  well known that
for length scales greater than $\delta$, the bare theory on the
$a$-dust can be replaced  by a renormalized theory defined on the
$\delta$-dust. The renormalized theory will typically be more
complicated, containing second, third, and $n^{th}$ neighbor
couplings.

Generally, the dimensionless  form of the renormalized theory will
depend on $\delta$ in just such a way that physics  at longer
scales is exactly the  same as it was in the original theory. The
dimensionless parameters will flow as the  reference scale is
changed.

If there is an infrared fixed-point, and if the bare theory is in
the basin of attraction of the fixed-point, then  as $\delta$
becomes much larger than $a$, the dimensionless parameters will
run to their fixed-point values. In that case the continuum limit
($a \to 0$) will be a conformal  field theory with $SO(3,1)$
invariance.

  These  things hold  in the holographic theory of  \ads. The infrared scale is
  provided by the geometry of the boundary, namely, a  sphere. The compact nature of the
  boundary gives the theory an energy gap.

  The bare scale is not so important in ADS/CFT. One might as well take the continuum
  limit from the start. On the other hand the running renormalization scale is all-important. It defines the
  holographically generated radial dimension of space. A useful slogan is  ``Motion
  along the $R$
direction is the same as renormalization-group flow." The reference scale $\delta$ is
  related to $R_0$ by $ \delta = e^{-R_0}$.

  More generally, we can allow the
  renormalization scale to depend on position along the boundary:
    \be
  \delta(\Omega) = e^{-R_0(\Omega)}.
  \label{delta-R relation}
  \ee

    Now to the second preliminary---some observations about Liouville Theory.
Again,   a concrete model is helpful. Liouville  theory
is closely connected with the theory of dense, planar, ``fishnet"
 diagrams \cite{holger} such as those which appear in large
N gauge theories, and matrix models \cite{hooft}\cite{shenker
douglas}\cite{migdal}.

We assume the fishnet has the topology of a sphere, i.e., it can be drawn on a sphere with no crossing of lines, and that it has a very large number of vertices. Following \cite{holger}, we draw the diagram so that it is locally isotropic. If it is not locally isotropic it may be made so by shifting the points around. Once the diagram is isotropic, the remaining freedom in drawing the fishnet is  conformal transformations.

The fishnet plays  the role  of the bare
lattice in  the previous discussion, but now it's dynamical---we
sum over all fishnet diagrams, assuming only that the spacing (on
the 2-sphere) is everywhere much smaller than the  sphere size, $L$.
As before, we call the  angular spacing between neighboring vertices
on the  sphere, $a(\Omega)$.

Each fishnet defines a metric on  the sphere.  Let $d\alpha$ be a
small angular interval (measured in radians). The fishnet-metric
is defined by
\be 
ds^2 = {d\alpha^2 \over a(\Omega)^2 }.
\label{deltafishmetric} 
\ee

We again introduce a reference scale $\delta$. It can also be
a fishnet, but now it is fixed, its vertices nailed down, not to
be integrated over. We continue to assume that $\delta $ satisfies
the inequalities, $a(\Omega) << \delta(\Omega)<<L$, but otherwise
it is arbitrary. The $\delta$-metric is  defined by 
\be
ds_{\delta}^2 = {d\alpha^2 \over \delta(\Omega)^2 }.
\label{deltametric} 
\ee

It should be clear that the metrics in (\ref{deltafishmetric}) and (\ref{deltametric}) are the same as the reference and real metrics in (\ref{mets}).
We can now define the  Liouville field $U$. All it is is the ratio
of the reference and fishnet
 scales:
\be e^U \equiv {\delta / a}. \label{liou} \ee Using (\ref{liou}) \
together with $\delta = e^{-R}$, and $ds = {d\alpha \over a}$, we
see that $U$ is also given by the relation, \be ds = d\alpha \
e^{(R_0+U)}. \label{liou metric} \ee

In (\ref{liou metric}) both $R_0$ and $U$ are functions of
location on $\Omega_2$, but only $U$ is dynamical, i.e., to be
integrated over.

\subsection{Liouville in the Hat}

With that in mind, we return to cosmic holography, and consider
the metric on the regulated spatial boundary of FRW, $\Sigma_0$.
In the absence of fluctuations it is
$$
ds^2 = e^{2R_0} e^{2T} d^2\Omega_2.
$$

In general relativity it is natural to allow both $R_0 $ and $T$
to vary over the sphere, so that \be ds^2 = e^{2R_0(\Omega_2)}
e^{2T(\Omega_2)} d^2\Omega_2 \label{metric T and R} \ee

The parallel between (\ref{liou metric}) and (\ref{metric T and R})
is obvious. Exactly as we might have expected from the \wdw \
interpretation, the Liouville field, $U$, may be identified with
time $T$,  when both are large, \be U \approx T \label{u
equals T}. \ee

The connection between $T$ and $U$ is ambiguous when $U$ becomes small. In that limit
the fishnet diagrams become very sparse and any detailed identification with the continuous variable $T$ breaks down.

To summarize, let's list a number of correspondences:
  \bea
\delta &\leftrightarrow& %\mu^{-1} \leftrightarrow
e^{-R},  \cr \lambda
&\leftrightarrow& e^{-T}, \cr a &\leftrightarrow& e^{-T^+} =
e^{-(T+R)}. \label{correspondences}
  \eea
  
  We also recall the correspondences with the real and reference metric:
\bea
ds^2&=&{d\alpha^2 \over a(\Omega)^2 }, \nonumber\\
ds_{\delta}^2 &=& {d\alpha^2 \over \delta(\Omega)^2 }.
\label{realanddeltametric}
\eea

One other point about Liouville Theory: the density of vertices of
a fishnet is normally varied by changing the weight assigned to
vertices. When the fishnet is a Feynman diagram the weight is a
coupling constant $g$. It is well known that the coupling constant
and Liouville \cc are alternate descriptions of the same thing.
Either can be used to vary the average vertex density---increasing
it either by increasing $g$ or decreasing $\lambda$.  The very
dense fishnets correspond to large $U$ and therefore large FRW
time, whereas very sparse diagrams dominate the  early Planckian
era.

\subsection{RG-covariant and RG-invariant Objects in Quantum Field Theory}

There are two kinds of objects\footnote{In an earlier version of this paper the terms ``proactive" and ``reactive" were used instead of RG-covariant and RG-invariant.} in Wilsonian renormalization \cite{wilson} that
correspond quite closely to the terms $G_1$ and $G_2$ that we have
found in   section 5. We call them \it``RG-covariant" \rm and
\it``RG-invariant" \rm. RG-covariant quantities
depend on the arbitrary reference scale.
They are not directly measurable quantities.
 The best example is the exact Wilsonian
action, defined at a specific reference scale. The form
of RG-covariant  quantities depends on that reference scale,
 and so does the value of their matrix elements; indeed
  their form  varies with $\delta$  in such a way as to
keep the physics fixed at longer distances.

By contrast, RG-invariant objects  are observables whose value does not
depend  on the reference scale. They do depend on the
``bare" cutoff scale $a$ through wave function renormalization
constants, which typically tend to zero as $a \to 0$. The wave
function renormalization constants are usually stripped off when
defining  quantum fields but we will find it more illuminating to
keep them.

The distinction between these two kinds of objects is subtle, and
is perhaps best expressed in Polchinski's version of the exact
Wilsonian renormalization group \cite{polchinski}. In that scheme,
at every scale there is a renormalized description in terms of
local defining fields $\phi(x)$, but the RG-covariant action grows
increasingly complicated as the reference scale is lowered.

Consider the exact effective action defined at reference scale
$\delta$. It is given by an infinite expansion of the form \be
L_{W}(\delta) = \sum_{\Delta = 2}^{\infty} g_{\Delta}
{\cal{O}}_{\Delta}, \label{wilsonian} \ee where
${\cal{O}}_{\Delta}$ are a set of operators of dimension $\Delta$,
and $ g_{\Delta}$ are dimensional coupling constants. The
renormalization flow is  expressed in terms of  the dimensionless
coupling constants, \be \tilde{g}_{\Delta} = g_{\Delta}
\delta^{(2-\Delta )}.
 \label{dimensionless couplings}
\ee The $\tilde{g}$ satisfy RG equations,
 \be
 {d\tilde{g}\over d \log{\delta}   } = - \beta(\tilde{g}).
 \label{rg equations}
 \ee

If the theory flows to a fixed point, in that limit the $\tilde{g}$ become constant.
 Thus the dimensional constants $g_{\Delta}$ in the
Lagrangian will grow with $\delta$. Normalizing them at the bare
scale $a$, in the fixed-point case we get 
\be g_{\Delta}(\delta) = g_{\Delta}(a) \
\left\{{\delta  \over a}\right\}^{(\Delta -2)}, \label{fixedpoint}
\ee
\be L_{W}(\delta) = \sum_{\Delta = 2}^{\infty} {\cal{O}}_{\Delta}
\left\{{\delta \over a}\right\}^{(\Delta -2)}. \label{wilsonian
expansion} \ee

Now consider the two point function of the effective action, $\<
L_{W}(\delta) L_{W}(\delta) \>$, evaluated at  distance scale
$L >> \delta$
\be
\< L_{W}(\delta) L_{W}(\delta)  \> =
\sum_{\Delta = 2}^{\infty}
 \< {\cal{O}}_{\Delta}
{\cal{O}}_{\Delta} \> \left({\delta \over a}\right)^{2(\Delta -2)}.
\label{LL}
\ee

Suppose the theory is defined on a sphere of radius $L$ and we
are interested in the correlator $\< L_{W}(\delta) L_{W}(\delta)
\>$ between points separated by angle $\alpha.$
The factor $\< {\cal{O}}_{\Delta}
{\cal{O}}_{\Delta} \>$ is the two-point function of a field of dimension $\Delta$,
in a theory on
the sphere of size $L$
with an ultraviolet cutoff at the reference scale $\delta$. Accordingly it has the form
\be
\< {\cal{O}}_{\Delta}
{\cal{O}}_{\Delta} \> = \left({\delta \over L}\right)^{2\Delta } (1-\cos
{\alpha})^{-\Delta}
\label{OO}
\ee
where the two factors of $\left({\delta \over L}\right)^{\Delta }$ are the ultraviolet-sensitive
wave function renormalization constants. The final result is
 \be \< L_{W}(\delta)
L_{W}(\delta)  \> =
  \sum_{\Delta = 2}^{\infty} C_{\Delta} \left({\delta \over a}\right)^{2(\Delta -2)}
\left({\delta \over L}\right)^{2\Delta } (1-\cos
{\alpha})^{-\Delta}.
 \label{wilsonwilson}
 \ee
 Note the
dependence of (\ref{wilsonwilson}) on the arbitrary reference
scale $\delta$. That dependence is typical of RG-covariant
quantities.

Now consider a RG-invariant quantity such as a fundamental  field,  a
derivative of such a field, or a local product of fields and
derivatives.   Their matrix elements at distance scale $L$ will be
independent of the reference scale (although it will depend on the
bare cutoff $a$) and be of order,
\be
\langle \phi \phi \rangle
\sim \left({a \over L}\right)^{2\Delta_{\phi}} (1-\cos
{\alpha})^{-\Delta_{\phi}}
\ee where
$\Delta_{\phi}$ is the operator dimension of $\phi$.  Thus we see
two distinct behaviors  for the scaling of correlation functions:
\be \left({\delta \over a}\right)^{2(\Delta -2)} \left({\delta
\over L}\right)^{2\Delta } \ \ \ \ \rm RG-covariant \label{RG-covariant}
\ee
 and
 \be
\left({a \over L}\right)^{2\Delta_{\phi}}  \ \ \ \ \rm RG-invariant.
\label{RG-invariant}
 \ee
The formulas are more complicated away from a fixed point but the
principles are the same.

We note that the effective action is not the only RG-covariant
object. The energy-momentum tensor  and various currents computed
from the effective action will also be RG-covariant. As we will see
these two behaviors---RG-covariant and RG-invariant---exactly correspond to
the dependence in (\ref{G1 in LC coords})  and (\ref{G2}).

Now we are finally ready to complete the discussion about the
relation between  the correlators of Section 5 and
RG-covariant/invariant operators. Begin by noting that in ADS/CFT, the
minimally coupled massless (bulk) scalar is the dilaton, and its
associated boundary field is the Lagrangian density. It may seem
puzzling that in the present case, an entire infinite tower of
operators seems to replace, what in ADS/CFT is a single operator.
In the case of the metric fluctuations a similar tower replaces
the energy-momentum tensor. The puzzle may be stated another way.
The FRW geometry consists of an infinite number of Euclidean ADS
time slices. At what time (or what 2D cosmological constant)
should we evaluate the boundary limits of the metric fluctuations,
in order to define the energy-momentum tensor? As we will see, a
parallel ambiguity exists in Liouville theory.

Return now, to the three scales of Liouville Theory: the infrared
scale $L$, the reference scale $\delta$, and  the fishnet scale
$a$, with $L>> \delta >> a$. It is natural to assume that the
basic theory is defined at the bare  fishnet scale $a$  by some
collection of degrees of freedom at each lattice site, and also
specific nearest-neighbor couplings---the latter insuring locality.
Now imagine a Wilsonian integration of all degrees of freedom on
scales between the fishnet scale and the reference scale,
including the fishnet structure itself. The result will be a
RG-covariant effective action of the type we described in equation
(\ref{wilsonian}). Moreover the correlation function of $L_{eff}$
will have the form (\ref{wilsonwilson}). But now, making the
identifications (\ref{delta-R relation}) and \be {\delta \over a}
= e^U = e^T, \label{identification} \ee we see that equation
(\ref{RG-covariant}) for  RG-covariant scaling becomes (for each
operator in the product) \be e^{({\Delta-2})T}e^{-\Delta R}.
\label{strength} \ee This is in precise agreement with the
coefficients in the expansion (\ref{G1}).

Similarly the RG-invariant
scaling (\ref{RG-invariant}),  $ e^{-\Delta T^+}$, is in agreement with
the properties of $G_2$.

What happens to the RG-covariant objects if we approach \sig \ by
sending $T^+ \to \infty$ and $T^-  \to - \infty$? In this limit only
the dimension-two term survives:  exactly what we would expect if
the matter action ran toward a fixed point. All of the same things
hold true for the tensor fluctuations. Before the limit $T^- \to
-\infty$, the energy-momentum tensor consists of an infinite
number of higher dimension operators but in the limit, all  tend
to zero except for the dimension two term.

It should be observed that the higher dimension contributions to
$G_1 \left\{^{i k}_{j l} \right\}$ are not transverse in the
two-dimensional sense. This is to be expected: before the limit is
taken, the Liouville field does not decouple from the matter
field, and the matter energy-momentum is not separately conserved.
But if the matter theory is at a fixed point, i.e., scale
invariant, the Liouville and matter do decouple and the matter
energy-momentum should be conserved. Thus, in the limit in which
the dimension-two term dominates, it should be (and is)
transverse-traceless.

The physical reason why the fluctuations of the Liouville field decouples at late time deserves some comment. As we have emphasized, the reason that the boundary geometry is dynamical is the asymptotic warmness of the FRW background as $R \to \infty$ at fixed time. But unlike de Sitter space, FRW cosmology becomes cold at late time. Thus it makes sense that the Liouville field decouples as $T \to \infty$.

RG flow is usually thought of in terms of a single independent
flow-parameter. In some versions it's the logarithm of the bare
cutoff scale, and in other formulations it's the log of the
renormalization scale. In the conventional ADS/CFT framework, $R$
can play either role. One can imagine a bare cutoff at some large
$R_0$ or one can push the bare cutoff to infinity and think of $R$
as a running renormalization scale.

However, for our purposes, it is better to keep track of both
scales. One can either think of a one-dimensional (logarithmic)
axis---we can call it the ``Wilson line"---extending from the
infrared scale to the fishnet scale $a$, or a two dimensional
$R,T$ plane. In either case
 the effective action as a function of two independent
variables. Figure 10 shows a sketch of the Wilson line and the two
dimensional plane representing the two directions $R$, and $T$.
 \bigskip
 \begin{figure}
\begin{center}
\includegraphics[width=20cm]{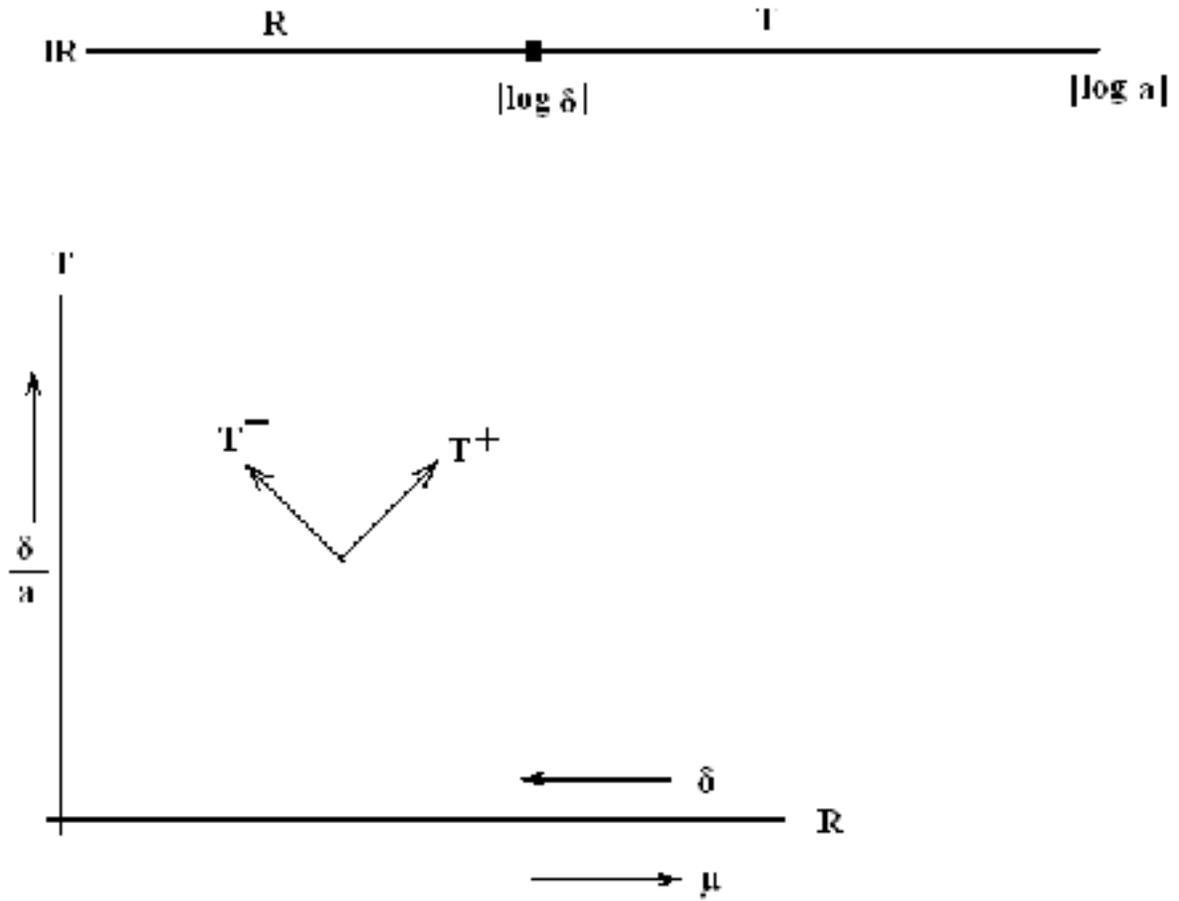}
\caption{The Wilson line of scales, and the two dimensional $R,T$
plane.} \label{7}
\end{center}
\end{figure}

The two independent parameters can be chosen to be $a$ and
$\delta$, or equivalently $R$ and $T$. Yet another choice is to
work in momentum space. The reference energy-scale is usually
called $\mu$, \be \mu = e^R \label{mu = e^R}. \ee And in the case
of negative central charge,
 the two-dimensional cosmological constant $\lambda$ can replace $T$.

In this light, it  is extremely interesting that the distinction
between RG-covariant and RG-invariant  scaling, corresponds to motion
along the two light-like directions $T^-$ and $T^+$ as depicted in
Figure 11.

 \begin{figure}
\begin{center}
\includegraphics[width=20cm]{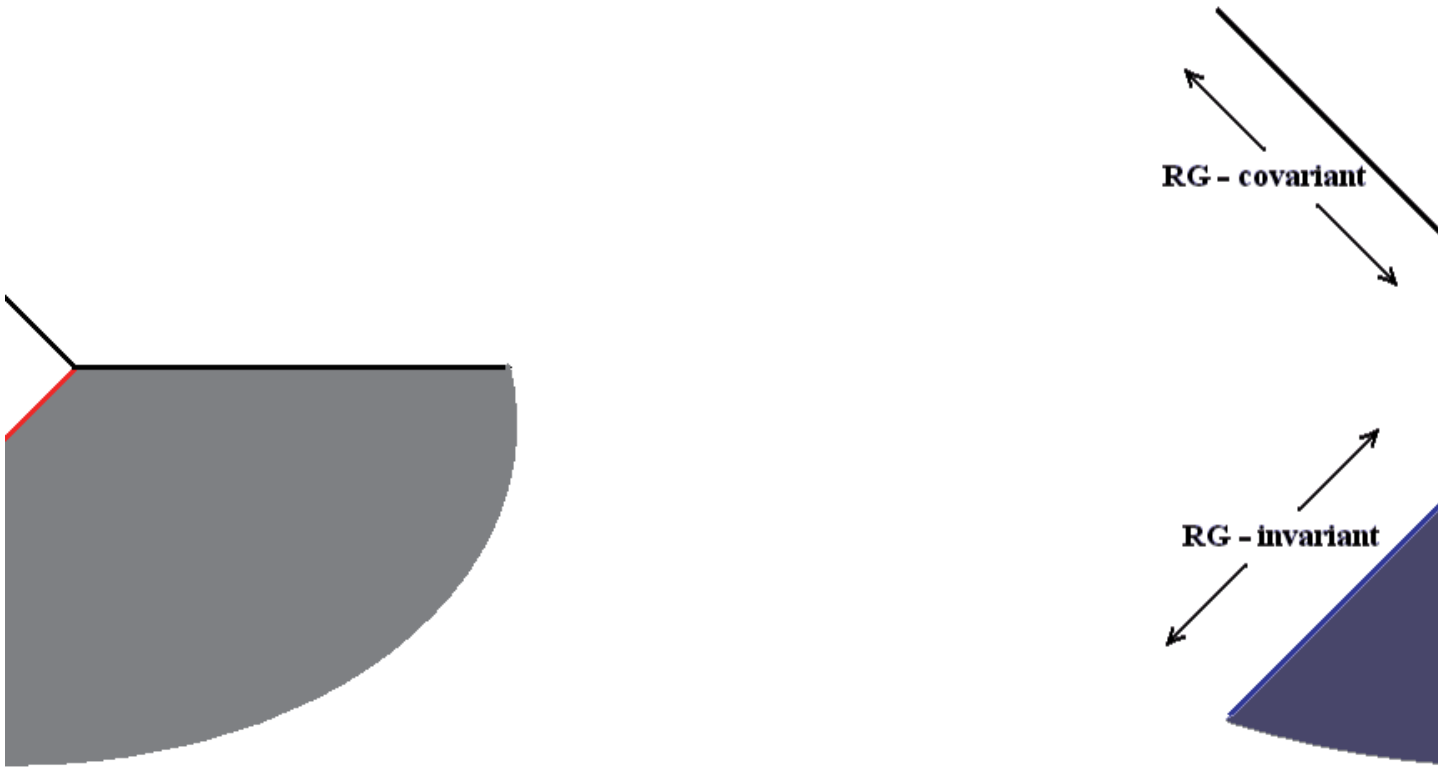}
\caption{RG-covariant and RG-invariant  quantities scale with the two
light-like directions $T^-$ and $T^+$.} \label{8}
\end{center}
\end{figure}

It may not be obvious  why the bulk fields should
correspond to RG-covariant and RG-invariant  boundary fields in the way
that they do. The solutions to the wave equation
in the bulk are generally sums of two types of modes, \bea \chi_-
&\to&   g_-(\Omega_2) \ F_-(T^-) \ e^{-2T}, \cr \chi_+ &\to&
g_+(\Omega_2) \ F_+(T^+). \label{chi plus and minus} \eea

Some insight can be obtained from Fig 8. The lower (red) light-like line represents the
initial condition of the FRW universe at $T=-\infty$.
In that limit the RG-covariant contributions
to the correlation functions tend to zero, and only the RG-invariant terms
are present. Thus it makes sense to think of the RG-invariant quantities as
inputs to the RG flow---in other words as the bare action. Given the input
on the red initial surface, the bulk field equations can be solved to
determine the output on the hat. Note that on the hat, the RG-invariant
quantities tend to zero, leaving only the RG-covariant.

This duality between the renormalization properties of Liouville theory and the  cosmological
coordinates $R$ and $T$ is very remarkable and clearly deserves more attention.
It is important to understand that the duality between FRW cosmology and Liouville 2D
gravity does not only involve the continuum fixed point theory. As long as $T$ is finite
the theory has some memory of the bare theory. Cosmology in the hat consists of an entire
RG flow from some bare theory to a final fixed point that governs late times.
It's only in the limit $T \to \infty$
that the
theory flows to the fixed point and loses memory
of the bare details. We will come back to this
point in section 8 when we discuss the Garriga, Guth, Vilenkin \cite{garriga}
``Persistence of Memory" phenomenon.

\setcounter{equation}{0}
  \section{ Scaling and the Census Taker}
  \subsection{Moments}

 Now we come to the  connection between the scaling
behavior of two-dimensional quantum field theory and the
observations of a Census Taker as he moves toward the Census
Bureau.
  In order to better understand the connection between the cutoff-scale  and $T^+$, let's
return to the similar connection between cutoff and the coordinate
$R$ in \ads.  We normalize the ADS radius of curvature to be $1$;
with that normalization, the Planck area is given by $1/c$ where
$c$ is the central charge.

  Consider the proper distance between points $1$ and $2$ given by (\ref{distance on H}).
The relation $l=R_1 +R_2 + \log{(1-\cos{\alpha})}$ is approximate,
valid when $l$ and $R_{1,2}$ are all large. When $l \sim 1$ or
equivalently, when
  \be
  \alpha^2 \sim e^{-(R_1 + R_2)}
  \label{cutoff angle large}
  \ee
equation (\ref{distance on H}) breaks down. For angles smaller
than (\ref{cutoff angle large}) the distance in \ads \ behaves
like \be l \sim e^{R}\alpha.
 \label{l small}
\ee Thus a typical correlation function will behave as a power of
$(1-\cos{\alpha})$ down to angular distances of order (\ref{cutoff
angle large}) and then fall quickly to zero.

The angular cutoff in \ads \ has a simple meaning. The solid angle
corresponding to the cutoff  is of order $e^{-2R}$ while the area
of the regulated boundary is $e^{+2R}$. Thus, metrically, the
cutoff area is of order unity. This means that in Planck units,
the cutoff area is the central charge $c$ of the boundary \cft.

%Now consider the cutoff angle implied by (\ref{massless 2D
%correlator with variable cutoff}) and (\ref{cutoff}). 
Now consider the cutoff angle in the hat. Equations 
(\ref{massless 2D correlator with variable cutoff}) and
(\ref{cutoff}) imply the UV cutoff is of order
%. 
%By an argument parallel to the one above, the cutoff angle becomes
 \be
  \alpha^2 \sim e^{-(T^+_1 + T^+_2)}.
  \label{cutoff angle in T+}
  \ee
Once again this corresponds to a small patch of proper area (on $\Sigma_0$, the
regulated boundary) which is time-independent, and in Planck
units,   of order the central charge. It is also equal to the area 
on the horizon in the ancestor vacuum. The degrees of freedom which
describes this patch will possibly be matrices as in the usual ADS/CFT 
correspondence.

%A region of this size will be described 
%by a finite number of degrees of freedom whose number is given 
%the central charge. These degrees of freedom will possibly be
%matrices as in the usual ADS/CFT correspondence. 

Consider the Census Taker  looking back  from some late time
$T_{CT}$.   For convenience we place the CT at $R=0$. His backward
light-cone is the surface \be T + R = T_{CT} \label{census T+}.
\ee The CT can never quite see \sig. Instead he sees the regulated
surfaces corresponding to a fixed proper cutoff  (Figure 4). 
The later the CT observes, the smaller the  angular structure that he
can resolve on the boundary. This is another example of the UV/IR
connection, this time in a cosmological setting.

Let's consider a specific example of a possible observation. The
massless scalar field $\chi$ of section 5 has an asymptotic limit
on \sig \ that defines the dimension zero field  $\chi(\Omega)$.
Moments of $\chi$ can be defined by integrating it with spherical
harmonics\footnote{We are using $l$ for geodesic distance on 
${\cal H}_3$, and $\ell$ for angular momentum quantum number.}, 
\be \chi_{\ell m} = \int \chi(\Omega) Y_{\ell m}(\Omega)d^2
\Omega. \label{moments} \ee It is worth recalling that in \ads \
the corresponding moments would all vanish because the
normalizable modes of $\chi$ all vanish exponentially as $R \to
\infty$. The possibility of non-vanishing moments is due entirely
to the asymptotic warmness of open FRW.

We can easily calculate\footnote{L.S. is indebted to Ben Freivogel
for explaining equation (\ref{mean square moment}).} the
mean square value of $\chi_{\ell m}$ (It is independent of $m$). \be
\< \chi_{\ell}^2 \> = H^{2}\int d(\cos\alpha)\ \log(1-\cos \alpha)
P_\ell(\cos \alpha) \sim
H^{2}{1 \over \ell(\ell+1)}. \label{mean square moment} \ee It is evident that
at a fixed Census Taker time $T_{CT}$, the angular resolution is
limited  by (\ref{cutoff angle in T+}). Correspondingly, the
largest moment that the CT can resolve corresponds to \be \ell_{max}\sim 
e^{T_{CT}}. \label{lmax} \ee Thus we arrive at the following
picture: the Census Taker can look back toward \sig \ but at any
given time his angular resolution is limited by (\ref{cutoff angle
in T+}) and (\ref{lmax}). As time goes on more and more moments
come into view. Once they are measured they are frozen and cannot
change. In other words the moments evolve from being unknown
quantum variables, with a gaussian probability distribution, to
classical boundary conditions that explicitly break rotation
symmetry (and therefore conformal symmetry). One sees from
(\ref{mean square moment}) that the  symmetry breaking is
dominated by the low moments.

This phenomenon never occurs in an undiluted form.
Realistically speaking, we don't expect massless scalars in the
non-supersymmetric ancestor. In Section 5 we discussed the effect
of a small mass term, in the ancestor vacuum, on the correlation
functions of $\chi$.
 The result of such a mass term is a
shift of the leading  dimension from $0$ to $\mu$. This has an
effect on the moments. The correlation function becomes \be
H^{2}e^{-\mu T_1^+} e^{-\mu T_2^+} (1 - \cos{\alpha} )^{-\mu}.
\label{shifted dimension} \ee and the moments take the form
 \be
\< \chi_{\ell}^2 \> = H^{2}e^{-2\mu T_{CT}} \int d(\cos\alpha)\ 
(1-\cos \alpha)^{\mu} \ P_\ell(\cos \alpha). 
\label{shifted square moment} \ee The functional
form of the $\ell$ dependence changes a bit, favoring  higher $\ell$,
but more importantly,
 the observable  effects  decrease like $e^{-2{\mu T_{CT}}}$. Thus as $T_{CT}$ advances,
the asymmetry on the Census Taker's sky decreases exponentially with conformal
time. Equivalently it decreases as a power of proper time along
the CT's world-line.

\subsection{ Homogeneity Breakdown}

Homogeneity in an infinite  FRW universe is generally taken for
granted, but before questioning homogeneity we should know exactly
what it means. Consider some three-dimensional scalar quantity
such as energy density, temperature, or the scalar field $\chi$.
Obviously the universe is not uniform on  small scales, so in
order to define homogeneity in a useful way we need to average
$\chi$ over some suitable volume. Thus at each point $X$ of space,
we integrate   $\chi$ over a sphere of radius $r$ and then divide
by the volume of the sphere. For a mathematically exact notion of
homogeneity the size of the sphere must tend to infinity. The
definition of the average of $\chi$ at the point $X$ is \be
\overline{\chi(X)} = \lim_{r \to \infty}{ \int \chi d^3x \over
V_r}.
  \label{average}
  \ee
 \begin{figure}
\begin{center}
\includegraphics[width=20cm]{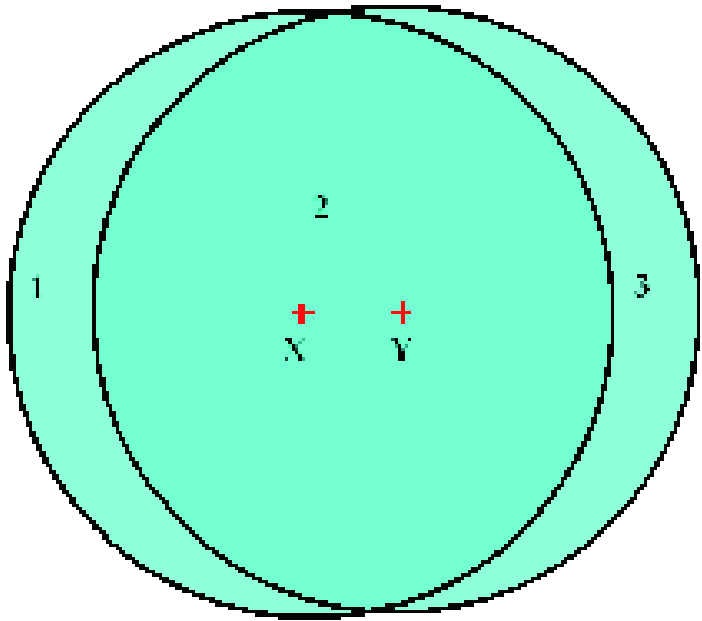}
\caption{Two large spheres centered at $X$ and $Y$.} \label{9}
\end{center}
\end{figure}
Now pick a second point $Y$ and construct $\overline{\chi(Y)}$.
The difference $\overline{\chi(X)}- \overline{\chi(Y)}$ should go
to zero as $r \to \infty$ if space is homogeneous. But as  the
spheres grow  larger than  the distance between $X$ and $Y$, they
eventually almost completely overlap. In Figure 12 we see that the
difference between $\overline{\chi(X)}$ and $\overline{\chi(Y)}$
is due to the two thin crescent-shaped regions, 1 and 3. It seems
evident that the overwhelming bulk of the contributions to
$\overline{\chi(X)}, \  \overline{\chi(Y)}$ come from the
central region 3, which occupies almost the whole figure. The
conclusion seems to be that the averages, if they exist at all,
must be independent of position. Homogeneity while true, is
trivial.

This is correct in flat space, but surprisingly it can break down
in hyperbolic space\footnote{L.S. is grateful to Larry Guth for
explaining this phenomenon, and to Alan Guth for emphasizing its
importance in cosmology.}. The reason is quite simple: despite
appearances the volume of regions 1 and 3 grow just as rapidly as
the volume of 2. The ratio of the volumes is of order \be {V_1
\over V_2} = {V_3 \over V_2} \sim  { l \over R_{curvature} } \ee
(where $l$ is the distance between $X$ and $Y$)
and remains finite as $r \to \infty$.

To be more precise we observe that \bea \overline{\chi(X)} &=& {
\int_1 \chi + \int_2 \chi  \over V_1 + V_2 } \cr
\overline{\chi(Y)} &=& { \int_3 \chi + \int_2 \chi  \over V_3 +
V_2 } \label{averages X,Y} \eea and that the difference
$\overline{\chi(X)}- \overline{\chi(Y)}$ is given by
\be
\overline{\chi(X)}- \overline{\chi(Y)}= { \int_1 \chi \over V_1 +
V_2 }- { \int_3 \chi \over V_3 + V_2 }
\label{difference}
\ee
which, in the limit $r \to \infty$ is easily seen to be
proportional to the  dipole-moment of the boundary theory,
 \be
 \overline{\chi(X)}
-\overline{\chi(Y)} =l  \int \chi(\Omega) \cos \theta d^2 \Omega =
l \  \chi_{1,0}. \ee

Since, as we have already seen for the case $\mu=0$, the mean
square fluctuation in the moments does not go to zero with
distance, it is also true that average value of
$|\overline{\chi(X)} -\overline{\chi(Y)}|^2$ will be nonzero. In
fact it grows with separation.

However there is no reason to believe that a dimension zero scalar
exists. Moduli, for example, are expected to be massive in the
Ancestor, and this shifts the dimension of the corresponding
boundary field. In the case in which the field $\chi$ has
dimension $\mu$, the effect (non-zero rms average of moments )
persists in a somewhat diluted form. If a renormalized field is
defined by  stripping off the wave function normalization
constants, $\exp{(-\mu T^+)}$, the squared moments still have
finite expectation values and break the symmetry. However, from an
observational point of view there is no reason
to remove these factors. Thus it seems that as the Census Taker
time tends to infinity, the observable asymmetry will decrease
like $\exp{(-2\mu T_{CT}})$.

\setcounter{equation}{0}
  \section{Bubble Collisions and Other Matters}

 By looking back toward \sig, the Census
Taker can see into bubbles of other vacua---bubbles that in the
past collided with his hatted vacuum.

As Guth and Weinberg recognized long ago \cite{guth weinberg}, a
single isolated bubble is infinitely unlikely. A typical
``pocket universe" will consist of a cluster of an unbounded
number of colliding bubbles, although if the nucleation rate is
small the collisions will in some sense be rare. To see why such
bubble clusters form it is sufficient to recognize why a single
bubble is infinitely improbable.
\begin{figure}
\begin{center}
\includegraphics[width=12cm]{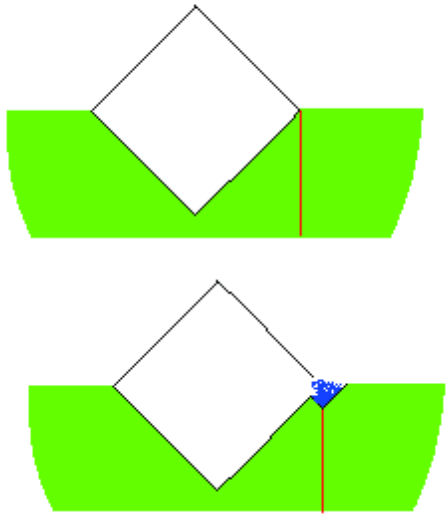}
\caption{The top figure represents a single nucleated bubble. The
red trajectory is a time-like curve of infinite length approaching
\sig. Because there is a constant nucleation rate along the curve,
it is inevitable that a second bubble will nucleate as in the
lower figure. The two bubbles will collide.} \label{12}
\end{center}
\end{figure}
In Figure 13 the main point is illustrated by drawing a time-like
trajectory that approaches \sig \ from within the Ancestor vacuum.
The trajectory has infinite proper length, and assuming that there
is a uniform nucleation rate, a second bubble will eventually
swallow the trajectory and collide with the original bubble.
Repeating this  process will produce an infinite bubble cluster.

More recently Garriga, Guth, and Vilenkin \cite{garriga} \ have
argued that the multiple bubble collisions must spontaneously
break the $SO(3,1)$ symmetry of a single bubble, and in the
process render the (pocket) universe inhomogeneous and
anisotropic. The breaking of symmetry in \cite{garriga} was
described, not as spontaneous breaking, but as explicit breaking
due to initial conditions. However, spontaneous symmetry breaking
$is$ the persistent  memory of a temporary explicit symmetry
breaking, if the memory does not fade with time. For example, a
small magnetic field in the very remote past will determine the
direction of an infinite ferromagnet for all future time.

The actual observability of bubble collisions depends on the
amount of slow-roll inflation that took place after tunneling.
Much more than 60 e-foldings \cite{maria} would probably wipe out any signal,
but the interest of this paper is conceptual.  We will take the
viewpoint that anything within the past light-cone of the Census
Taker is in principle observable.

In the last section we saw that perturbative infrared effects are
capable of breaking the $SO(3,1)$ symmetry, and it is an
interesting question what the relation between these two
mechanisms is. The production of a new bubble would seem to be a
non-perturbative  effect that adds to the perturbative symmetry
breaking effects of the previous section. Whether it adds
distinctly new effects that are absent in perturbation theory is
not obvious and may depend on the specific nature of the
collision. Let us classify some possibilities.

\subsection{Collisions with Identical Vacua}

The simplest situation is if the true-vacuum bubble collides with
another identical bubble, the two bubbles coalescing to form a
single bubble, as in the top of Figure 14.

 \begin{figure}
\begin{center}
\includegraphics[width=12cm]{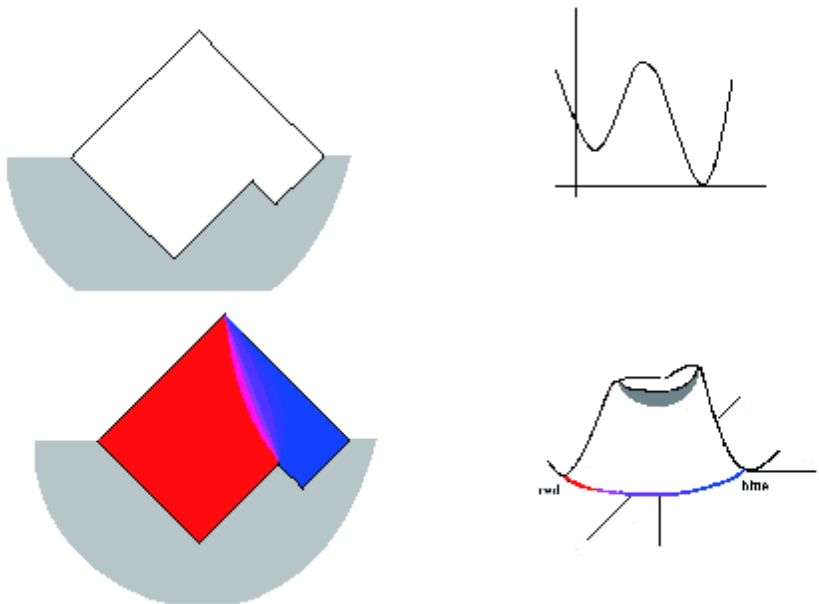}
\caption{In the top figure two identical bubbles collide. This
would be the only type of collision in a simple landscape with two
discrete minima---one of positive energy and one of zero energy. In
the lower figure a more complicated situation is depicted. In this
case the false vacuum $F$ can decay to two different true vacua,
``red"  and ``blue," each with vanishing energy. The two true
vacua  are connected by a flat direction, but CDL instantons only
lead to the red and blue points.} \label{11}
\end{center}
\end{figure}

 The surface \sig \  is defined by starting at the tip of the hat and tracking back along
light-like trajectories until they end---in this case at a false
vacuum labeled \bf F. \rm
 The collision is parameterized by the space-like separation between nucleation points.
Particles produced at the collision of the bubbles just add to the
particles that were produced by ordinary FRW evolution. The main
effect of such a collision is to create a very distorted boundary
geometry, if the nucleation points are far apart. When they are
close the double nucleation blends in smoothly with the single
bubble. These kind of collisions seem to be no different than the
perturbative disturbances caused by the non-normalizable mode of
the metric fluctuation. Garriga, Guth, and Vilenkin, compute that
the typical observer will see multipole moments on the sky, but as
we've seen, similar multipole moments  can also occur
perturbatively.

In the bottom half of Figure 14 we see another type of collision
in which the colliding bubbles correspond to two different true
vacua: red (r) and blue (b). But in this case red and blue are on
the same moduli-space, so that they are connected by a flat
direction\footnote{We assume that there is no symmetry along the flat
direction, and that there are only two tunneling paths from the false vacuum, one  to
red, and one to blue.}. Both vacua are included within the hat. In the bulk
space-time they bleed into each other, so that as one traverses a
space-like surface, blue gradually blends into purple and then
red.

On the other hand the surface \sig \ is sharply divided into blue
and red regions, as if by a one dimensional domain wall. This
seems to be  a new phenomenon that does not occur in perturbation
theory about either vacuum.

As an example, consider a case in which  a red  vacuum-nucleation
occurs first, and then much later a blue vacuum bubble nucleates.
In that case the blue patch on the boundary will be very small and
The Census Taker will see it occupying tiny angle on the sky. How
does the boundary field theorist interpret it? The best
description is probably as a small blue ``instanton" in a red vacuum\footnote{We are using the term ``instanton" very loosely. On the one hand, the occurrence of such blue patches is non-perturbative and exponentially suppressed by the decay rate in the bulk. But there does not seem to be any topological stability to the object.  }.
In both the bulk and boundary theory this is an exponentially
suppressed,  non-perturbative effect.

However, in a \cft \ the size of an instanton is a modulus that
must be integrated over. As the instanton grows the blue region
engulfs more and more of the boundary. Eventually the
configuration evolves to a blue 2D vacuum, with a tiny red
instanton. One can also think of the two configurations as the
observations of two different Census Takers at a large separation
from one  another. Which one of them is at the center, is
obviously ambiguous.

The same ambiguous separation into dominant vacuum, and small
instanton, can be seen another way. The nucleation sites of the
two bubbles are separated by a space-like interval. There is no
invariant meaning to say that one occurs before the other. A
element of the de Sitter symmetry group can interchange which
bubble nucleates early and which nucleates later.

Nevertheless, a given Census Taker will see a definite pattern on
the sky. One can always define the CT to be at the center of
things, and integrate over the relative size of the blue and red
regions. Or one can keep the size of the regions fixed---equal for
example---and integrate over the location of the CT.

From both the boundary field theory, and the bubble nucleation
viewpoints, the probability for any finite number of red-blue
patches is zero. Small red instantons will be sprinkled on every
blue patch and vice versa, until the boundary becomes a fractal.
The fractal dimensions are closely connected to operator
dimensions in the boundary theory. Moreover, exactly the same
pattern is expected from multiple bubble collisions.

But the Census taker has a finite angular resolution. He cannot
see angular features smaller than $\delta \alpha \sim
\exp{(-T_{CT})}$. Thus he will see a finite sprinkling of red and
blue dust on the sky. At $T_{CT}$ increases, the UV cutoff scale
tends to zero and the CT sees a  homogeneous``purple" fixed-point
theory.

The red and blue patches are reminiscent of the Ising spin system
(coupled to a Liouville field). As in that case, it makes sense to
average over small patches and define a continuous ``color field"
ranging from intense blue to intense red. It is interesting to ask
whether \sig \ would look isotropic, or whether there will be
finite multipole moments of the renormalized  color field (as
in the case of the $\chi$ field). The calculations of Garriga,
Guth, and Vilenkin suggest that multipole moments would be seen.
But unless for some reason there is a field of exactly zero
dimension, the observational signal should fade with Census Taker
time.

There are other types of collisions that seem to be fundamentally
different from the previous. Let us consider a model landscape
with three vacua---two false, $B$ and $W$ (Black and White); and
one true vacuum $T$. Let the the vacuum energy of   $B$ be bigger
than that of $W$, and also assume that the decays $B \to W$, $B
\to T$, and $W \to T$ are all possible. Let us also start in the
Black vacuum  and consider a transition to the True vacuum. The
result will be a hat bounded by \sig.

 \begin{figure}
\begin{center}
\includegraphics[width=12cm]{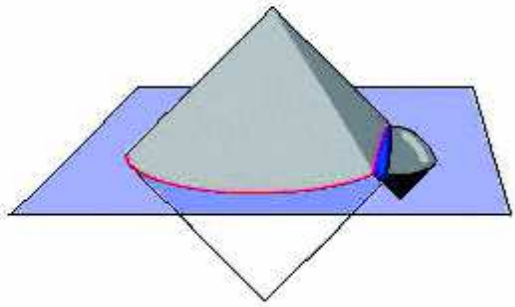}
\caption{A bubble of True vacuum forms in the Black false vacuum
and then collides with a bubble of White vacuum. The true vacuum
is bounded by a hat but the White vacuum terminates in a
space-like surface. Some generators of the hat intersect the Black
vacuum and some intersect the White. Thus \sig, shown as the red
curve, is composed of two regions. } \label{14}
\end{center}
\end{figure}
However, if a bubble of $W$ forms, it may collide with the $T$
bubble  as in Figure 15. The $W$ bubble does not end in a hat but
rather, on a space-like surface. By contrast, the true vacuum
bubble does end in a hat. The surface \sig \ is defined as always,
by following the light-like generators of the hat backward until
they enter the bulk---either Black or White---as  in Figure 15.

In this case a portion of the boundary  \sig \ butts up against
$B$, while another portion abuts $W$.
 In some
ways this situation is similar to the previous case where the
boundary was separated into red and blue regions, but there is no
analog of the gradual bleeding of vacua in the bulk. In the
previous case the Census Taker could  smoothly pass from red to
blue.  But  in the current example, the CT would have to crash
through a domain wall in order to pass from $T$ to $W$. Typically
this happens extremely fast, long before the CT could do any
observation. In fact if we define Census Takers by the condition
that they eventually reach the Census Bureau, then they simply
never enter $W$.

From the field theory point of view  this example leads to a
paradox. Naively, it seems that once a $W$ patch forms on \sig, a
$B$ region cannot form inside it. A constraint of this type on
field configurations would obviously violate the rules of quantum
field theory; topologically (on a sphere) there is no difference
between a small $W$ patch  in a $B$ background, and a small $B$
patch  in a $W$ background. Thus field configurations must exist
in which a $W$ region has smaller  Black spots inside it. There is
no way consistent with locality  to forbid bits of
$B$ in regions of $W$.

Fortunately the same conclusion is reached from the bulk point of
view. The rules of  tunneling transitions require that if the
transition $B \to W$ is possible, so must be the transition $W \to
B$, although the probability for the latter  would be  smaller (by
a large  density of states ratio). Thus one must expect $B$  to
invade regions of $W$.

As the Census Taker time advances he will see smaller and smaller
spots of each type. If one assumes that there are no operators of
dimension zero, then the pattern should fade into a homogeneous
average grey,  although under the conditions we described it will
be almost White.

The natural interpretation is that the boundary field theory has
two phases of different free energy, the  $B$ free energy being
larger than that of  $W$. The dominant configuration would be the
ones of lower free energy with occasional fluctuations to higher
free energy.

\subsection{The Persistence of Memory }

Returning
to Figure 13, one might ask why no bubble formed along the red
trajectory in the infinitely remote past. The authors of
\cite{garriga} \ argue that eternal inflation does not make sense
without an initial condition specifying a past surface on which no
bubbles had yet formed. That surface invariably breaks the O(3,1)
symmetry and distinguishes a ``preferred Census Taker" who is at
rest in the frame of the initial surface.  He alone sees an
isotropic sky whereas all the other Census Takers see non-zero
anisotropy. What's more the effect persists no matter how late the
nucleation takes place.

The ``Persistence of Memory" reported in \cite{garriga} \ had
nothing to do with whether or not the Census Taker's sees a fading
signal: Garriga, Guth, and Vilenkin were not speaking about Census
Taker time at all. They were referring to the fact that no matter
how long after the start of eternal inflation a bubble nucleates,
it  will remember the symmetry breaking imposed by the initial
conditions; not whether the signal fades with $T_{CT}$.

To be clear about this, consider two (proper) times, $t_N$ and $t_{CT}$. The first, $t_N$ is
the time after the initial condition at which the Census Taker's bubble nucleates. The
second, $t_{CT}$, is the  Census Taker time  measured from the nucleation event.
The persistence of memory refers to $t_N$. No matter how late the bubble nucleates, the
Census Taker will see some memory of the initial conditions at finite $t_{CT}$.

An entirely separate question is whether the symmetry breaking effects of the initial
condition fade away with $t_{CT}$ and if so how fast. The answer to
this is determined by the spectrum
of dimensions in the conformal field theory. If there are no dimension-zero operators
the effects will dilute with increasing Census Taker time.

Let us consider  how the Persistence of Memory  fits together with the
RG flow discussed earlier\footnote{These observations are based on
work with Steve Shenker.}. Begin by considering the behavior for
finite $\delta$ in the limit of small $a$. It is reasonable to
suppose that in integrating out the many scales between $a$ and
$\delta$, the theory will run to a fixed point. Now recall that
this is the limit of very large $T$. If in fact the theory has run
to a fixed point it will be conformally invariant. Thus we expect
that the symmetry $O(3,1)$ will be unbroken at very late time.

On the other hand consider the situation of $\delta/ a$ of order $1$.
The reference and bare scales are very close and there are few degrees
of freedom to integrated out. There is no reason why the
effective action should be near a fixed point. The implication is
that at very early time (recall, $\delta/ a = e^T$)  the physics
on a fixed time slice will not be  conformally invariant. Near
 the beginning of an RG flow the effective action is strongly
dependent on the bare theory. The implication of a breakdown of
conformal symmetry is that there is  no symmetry between Census
Takers at different locations in space. In such situations the
center of the (deformed) \ads \ is, indeed, special.
The GGV boundary condition at the onset
of eternal inflation is the same thing as the initial condition on
the RG flow. In other words, varying the GGV boundary condition is
no different from varying the bare fishnet theory.

Is it possible to tune the bare action so that the theory starts
out at the fixed point? If this were so, it would be an initial
condition that allowed exact conformal invariance for all time. Of
course it would involve an infinite amount of fine tuning and is
probably not reasonable. But there may be reasons to doubt that it
is possible altogether, even though in a conventional lattice
theory it is possible.

The difficulty is that the bare and renormalized theories are
fundamentally different. The bare theory is defined on a variable
fishnet whose connectivity is part of the dynamical degrees of
freedom. The renormalized theory is defined on the fixed reference
lattice. The average properties of the underlying dynamical
fishnet are replaced by conventional fields on the reference
lattice. Under these circumstances it is hard to imagine what it
would mean to tune the bare theory to an exact fixed point.

The example of the previous subsection involving two false vacua,
$B$ and $W$,  raises some interesting questions. First imagine
starting with GGV boundary conditions such that, on some past
space-like surface, the vacuum is pure Black and that a bubble of
true vacuum nucleates in that environment. Naively the boundary is
mostly black. That means that in the boundary
theory the free energy of Black must be lower than that of
White.

But we argued earlier that white instantons will eventually
fill \sig \ with an almost white, very light grey color, exactly
as if the initial GGV condition were White. That means that White
must have the lower potential energy. What then is the meaning of
the early dominance of $B$ from the 2D field theory viewpoint?

The point is that it is possible for  two rather different bare
actions to be in the same broad basin of attraction and flow to
the same fixed point. The case of Black GGV conditions corresponds
to a bare starting point (in the space of couplings) where the
potential of $B$ is lower than $W$. During the course of the flow
to the fixed point the potential changes so that at the
fixed-point $W$ has the lower energy.

On the other hand, White  GGV initial conditions corresponds to
starting the flow at a different bare point---perhaps closer to the
fixed point---where the potential of $W$ is lower.

This picture suggests a powerful principle. Start with the space
 of  two-dimensional actions,  which
is broad enough to contain a very large Landscape of 2D theories.
With enough fields and couplings the space could probably contain
everything. As  Wilson explained \cite{wilson}, the space divides itself
into basins of attraction. Each initial state of the universe is
described either as a GGV initial condition, or as a bare starting
point for an RG flow. The endpoints of these flows correspond to
the possible final states-the hats---that the Census Taker can end
up in.

We have not exhausted all the kinds of collisions that can
occur---in particular collisions with singular, negative \cc \
vacua. A particularly thorny situation results if there is a BPS
domain wall between the negative and zero CC bubbles, then as
shown by Freivogel, Horowitz, and Shenker \cite{FHS} the entire
hat may disappear in a catastrophic crunch. The meaning of this is unclear.

\subsection{Flattened Hats }

In a broad sense this paper is about phenomenology: the Census
Taker could be  us: If we lived in an ideal thin-wall hat we would
see, spread across the sky, correlation functions of a holographic
quantum field theory: We could measure the dimensions of operators
both by the time dependence of the received signals, and their
angular dependence: Bubble collisions would appear as patches
resembling instantons.

Unfortunately (or perhaps fortunately) we are insulated from these
effects by two forms of inflation---the slow-roll inflation that
took place shortly after bubble nucleation---and the current
accelerated expansion of the universe. The latter means that we
don't live in a true hatted geometry. Rather we live in a
flattened hat,  at least if we ignore the final decay to a
terminal vacuum.

 \begin{figure}
\begin{center}
\includegraphics[width=12cm]{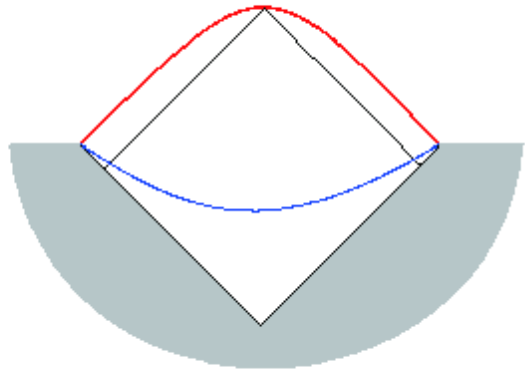}
\caption{If a CDL bubble leads to a vacuum with a small positive
cosmological constant, the hat is replaced by a rounded space-like
surface. The result is that no Census Taker can look back to
\sig.} \label{13}
\end{center}
\end{figure}

The  Penrose diagram  in Figure 16 shows an an Ancestor, with
large vacuum energy, decaying to a vacuum with a very small
cosmological constant. The important new feature is that the hat
is replaced by a space-like future infinity. Consider the Census
Taker's final observations as he arrives at the flattened hat. It
is obvious from  Figure 16 that he cannot look back to \sig. His
past light cone is at a finite value of $T^+$. Thus for each
time-slice $T$, there is  a maximum radial variable $R = R_0(T)$
within his ken, no matter how long he waits\cite{matt}. In other words there
is an unavoidable ultraviolet cutoff. It is  evident
that a final de Sitter bubble must be described by a theory with
no continuum limit; in other words not only a non-local theory,
but one with no ultraviolet completion\footnote{This suggests that 
de Sitter Space may have an intrinsic
imprecision. Indeed, as Seiberg has emphasized, the idea of a
metastable vacuum is imprecise, even in condensed matter physics
where they are common.  L.S. is grateful to Nathan Seiberg for
discussions on this point.}.

Another, perhaps more serious limitation,  is that  all of the memory of a past bubble nucleation
may, for observational purposes, be  erased by the slow-roll
inflation that took place shortly after the \cdl \ tunneling
event---unless it lasted for the minimum permitted number of
e-foldings \cite{maria}. In principle the effects are imprinted on
the sky, but in an exponentially diluted form.

Nevertheless, it may be interesting to explore the phenomenology of a limiting case in which the amount of slow roll inflation is very near the observational lower bound \cite{maria} and in which the cosmological constant is non-zero but arbitrarily small. The point of this exercise would be to get some idea of the possible effects of eternal inflation and how they are encoded in the FRW/CFT correspondence. In this paper we will only make a few simple observations. In order to keep track of the Ancestor \cc \ and the present \cc \ we will use the notation $H$ for the Ancestor Hubble constant, and $h$ for the current value.

The simplest way to compare the Census Taker with an observer in the later stages of a universe with a non-zero \cc \ is to simply cut off the RG flow when the area $H^{-2} \exp{(2T^+)}$ is equal to the horizon area in the late-time de Sitter vacuum, namely $h^{-2}$. Thus we simulate the effect of the event horizon by imposing a final value of $T^+$---call it $T^+_f$,
\be
e^{T^+_F} \sim {H \over h }.
\ee

As an example consider the situation described in Section 7. 

\bigskip

   \section{Acknowledgments}

 We are grateful to   Ben Freivogel,    Chen-Pin Yeh,
  Raphael Bousso,
 Larry Guth, Alan Guth,   Simeon Hellerman,  Matt Kleban,  Steve Shenker
  and Douglas Stanford for many insightful discussions. 
The work of YS is supported in part by MEXT 
Grant-in-Aid for Young Scientists (B) No.21740216.

%%%%%%%%%%%%%%%%%%%%%%%%%%%%%%%%%%%%%%%%%%%%%%%%%%%%%%%%%%%%%%
%%%%%%%%%%%%%%%%%%%%%%%%%%%%%%%%%%%%%%%%%%%%%%%%%%%%%%%%%%%%%%

\end{document}